\documentclass[aip,cha,reprint,floatfix,longbibliography]{revtex4-1}

\usepackage{amssymb}
\usepackage{amsmath}
\usepackage{amsfonts}
\usepackage{graphicx}
\usepackage{epstopdf}
\usepackage{epsfig}
\usepackage{xcolor}
\usepackage[normalem]{ulem} 
\usepackage{url}

\usepackage{lineno}
\renewcommand{\thesubsection}{\thesection.\Alph{subsection}}
\makeatletter
\def\p@subsection{}
\makeatother

\newcommand{\largefigure}{0.9\columnwidth}
\newcommand{\halffigure}{0.45\columnwidth}
\newcommand{\wb}{\omega_\text{b}}
\newcommand{\bre}{_\text{b}}
\newcommand{\wo}{\omega_0}

\newcommand{\wmax}{\omega_\text{max}}
\renewcommand{\wp}{\omega_{0,\text{phys}}}

\newcommand{\ket}[1]{|#1\rangle}
\newcommand{\bra}[1]{\langle #1|}
\newcommand{\braket}[2]{\langle #1 | #2\rangle}
\newcommand{\dyad}[2]{|#1 \rangle \langle  #2|}

\newcommand{\qt}{s}
\newcommand{\pt}{h}

\def\cn2{|c_n|^2}

\newcommand{\ii}{\textrm{i}}

\renewcommand{\d}{\textrm{d}}

\begin{document}


\title{Floquet breathers in a modulated nonlinear lattice}%
\author{Masayuki Kimura}
\affiliation{Department of Electrical and Electronic Engineering, Faculty of Science and Engineering\\Setsunan University,
17-8 Ikeda-Nakamachi, Neyagawa, Osaka 572-8508, Japan}
\author{Juan F. R. Archilla}
\email[Corresponding author:\,]{archilla@us.es}
\affiliation{Grupo de F\'{\i}sica No Lineal,  Universidad de
Sevilla, ETSI Inform\'{a}tica,
Avda Reina Mercedes s/n, 41012-Sevilla, Spain}
\author{Yusuke Doi}
\affiliation{Division of Mechanical Engineering, Graduate School of Engineering, The University of Osaka,
2-1 Yamadaoka, Suite, Osaka 565-0871, Japan}
\author{V\'ictor J. S\'anchez-Morcillo}
\affiliation{Universitat Polit\`ecnica de Val\`encia, Instituto de Investigaci\'on para la Gesti\'on Integrada de Zonas Costeras (IGIC),
Paranimf 1, 46730 Grao de Gandia, Val\`encia, Spain}

\begin{abstract}
In this work, we study a space-time modulated electro-mechanical system, consisting of an array of coupled cantilevers with their on-site potential provided by electromagnets driven by AC currents. Model equations are derived, and the effect of the modulation on the dispersion bands is examined. The theory of breather existence and stability is extended to include space-time modulation. We perform numerical simulations in a time-modulated system, showing three types of breather response depending on the driving frequency: (i) the modulation frequency is an integer multiple of the breather frequency or, in other words, this phenomenon corresponds to period doubling, tripling, etc.; (ii) the opposite, that is, the breather frequency is an integer multiple of the modulation frequency, corresponding to period-halving, etc. (iii) the breather and modulation frequencies are commensurate in a different form. We use for all of them the term {\em Floquet breathers} in analogy with Floquet solitons in photonic systems. As there is no dissipation, but periodic forcing, the energy is generally conserved\label{rev22-1B}  but only at discrete times.  There exists in this system a huge variety of breathers, either site-centered, symmetric and antisymmetric, bond centered, in-phase or in-quadrature with the modulation, and we analyze the evolution of stability of some of them as a function of the modulation frequency.
The construction of a similar system would be of interest to study the properties of dynamic metamaterials.
\end{abstract}
\keywords{
nonlinear waves, breathers, thermal equilibrium, localization, intrinsic localized modes (ILM)}
\pacs{63.20.Pw, 
  63.20.Ry,  
 05.45.-a,	
02.70.-c 
}

\maketitle

\newcommand{\fracc}[2]{\frac{\displaystyle #1}{\displaystyle #2}} 

{\bf We propose a space-time modulated system formed by an array of cantilevers where the on-site potential is provided by electromagnets fed with DC and AC currents. The modulation changes the phonon bands and the theory of breathers, or localized periodic solutions. We obtain breathers for many different frequencies, larger or smaller than the modulating one, but always commensurate.  We analyze in particular the stability dependence on the frequency for equal-period, doubled-period and halved-period breathers.  The results are of application for dynamic metamaterials.}

\section{Introduction}
\label{sec:intro}
Highly localized periodic vibrations, so-called intrinsic localized modes (ILM) or discrete breathers (DB) were first reported by Sievers and Takeno.\cite{sieverstakeno1988} They are ubiquitous in many nonlinear lattice-type systems.\cite{flachgorbach2005,flachgorbach2008}
ILMs have been shown to exist with different spatial patterns; the simplest one is the Sievers-Takeno  (ST) mode,\cite{sieverstakeno1988} where a single oscillator, called the center, has an amplitude much larger than its nearest neighbors, so the amplitude decreases rapidly with the distance to the center. Other mode is the Page (P) mode,\cite{page1990} where two neighboring oscillators have the same amplitude, either in-phase or anti-phase, being therefore a double ILM or breather.

An important issue regarding ILMs concerns their stability. The stability of ILMs is related to their spatial symmetry and the
Peierls-Nabarro potential barrier.\cite{kivshar-campbell1993}
The stability properties of these modes depend
on the class of lattice model considered. In lattices with nonlinear on-site
potentials, such as nonlinear Klein-Gordon (NKG) systems, rigorous results on existence and stability of DB
has been given.\cite{mackayaubry94}

The study of breather stability was considerably enhanced by Aubry's band theory, where the eigenvalues of the Newton operator, that is, the operator corresponding to the perturbation of a breather, allow for precise determination of the linear\label{rev22-3} stability and structural stability of breathers and multibreathers.\cite{aubry1997}

In this work, we are interested in breathers in a time-periodic nonlinear system. We will use the term {\em Floquet breathers}, similar to Floquet solitons used in photonic systems.\cite{mukherjee-science2020,mukherjee-rechtsman2023,kang-period-halving2025,parker-floquet-solitons2022,johansson-floquet-zigzag2025} Therefore, Floquet breathers are time-periodic, spatially localized vibrations in discrete, time-periodic nonlinear systems.  The breather period does not need to be the period of the system, but commensurate with it, as used in the last references. We will consider a specific time-periodic nonlinear system as described below.

Arrays of coupled cantilevers have played a relevant role in the study of nonlinear vibrations, in particular on energy localization,\cite{satohubbardsievers2006} and they are among the physical systems where breathers have been reported. They can be built at very different sizes, including nano-, micro- and macro-scales, and can be driven by different types of forces. They are part of important applications, such as actuators and sensors.

An experimental macroscopic array of cantilevers was introduced by Kimura and Hikihara in Ref.\,\onlinecite{kimura2009a}. The system was driven by a periodic force applied to the support of the cantilevers, appearing as an independent external (additive) force in the dynamical equations. The restoring force had two contributions: one provided by the elastic properties of the cantilever, and the other of magnetic nature, by fixing a magnet at the free end of the cantilever, and locating an electromagnet below it. The magnetic force due to permanent magnet-electromagnet interaction, can be tuned by changing an electric DC current, in this way, a tunable on-site potential was obtained.

The system resulted in breathers having different properties depending on the current magnitude and the frequency.
These and similar models have been used in the study of ILMs.\cite{satosievers2007,kimura2009b,kimura2016,chong2019,araki2024}

In this paper we present a variation of the model proposed in Ref.\,\onlinecite{kimura2009a}, incorporating, besides the static (DC) current, an oscillating (AC) contribution in the electromagnets. This brings about an on-site potential that it is tunable and periodic in time. Furthermore, introducing phase differences between neighboring cantilevers, an on-site potential with a space-time modulation is achieved. An experimental and mathematical model with a linear on-site potential and nonlinear coupling has been studied,\cite{chong2024}  finding $q$-gap breathers. In our case, the on-site potential is nonlinear and the coupling is linear.

The article is organized as follows: after the introduction in Sec.\,\ref{sec:intro}, Sec.\,\ref{sec:physical} presents the physical model and the derivation of the system of differential equations, both for the nonlinear system and its linear approximation corresponding to small oscillations. In Sec.\,\ref{sec:dispersionrelation}, the effect of space and time modulation on the dispersion bands is presented, while the theoretical deduction is located in Appendix\,\ref{app:dispersion}.
Sec.\,\ref{sec:theory_ILM} presents changes brought about by modulation in breather theory, breather obtention, and linear\label{rev22-3}
stability.
 Sec.\,\ref{sec:timecrystals},  presents breathers with a period that is an integer of the modulation period, and thus with a frequency smaller than the modulating one, and their properties are analyzed. The case of the breather frequency as a multiple of the modulation frequency is presented in Sec.\,\ref{sec:antitimecrystals}. Sec.\,\ref{sec:frequencydependence} considers also rational ratios of the breather and modulation frequencies, and the possibility of the breathers to be in quadrature with the modulation. A systematic study of the evolution of linear\label{rev22-3}
stability with the frequency is performed.  The final Sec.\,\ref{sec:conclusions} presents the concluding remarks.
Important but long explanations and mathematical derivations are included in three appendices in order not to interrupt the flow of the article focused on Floquet breathers.
Appendix\,\ref{app:thermal} deals with the concept of thermalization and numerical experiments that are used to check the theory. Appendix\,\ref{app:notation} presents concepts and notation used in some of the main text and in Appendix\,\ref{app:dispersion}, where the mathematical derivations of the dispersion relations are included.

\section{Physical system and dynamical equations}
\label{sec:physical}
In this section, we derive from physical laws the system dynamical equations. A sketch of the physical model is shown in Fig.\,\ref{fig_cantilevers}.
\setcounter{figure}{0}
\begin{figure}[htbp]
\begin{center}
\includegraphics[width=\largefigure]{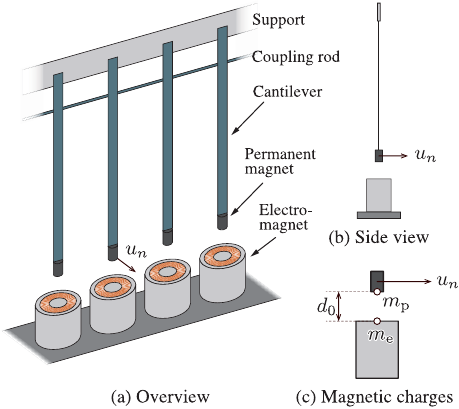}\end{center}
\caption{(a) System of cantilevers mechanically coupled by a rod. (b) Side view of a cantilever. A permanent magnet is attached to the tip of the cantilever. An electromagnet is placed below the
permanent magnet. (c) Definition of magnetic charges.
 }
\label{fig_cantilevers}
\end{figure}

The variable $u_n$ measures the horizontal deviation of the cantilever tip with respect to its equilibrium position. The vertical distance between the electromagnet and the cantilever tip is $y_n$, therefore, the position vector of the tip with respect to the electromagnet is $\vec r_n=u_n\hat{e}_x+y_n\hat{e}_y$, with equilibrium position $\vec r_{n,0}=d_0\hat{e}_y$, and modulus $r_n=\sqrt{u_n^2+y_n^2}$. Within the formulation of magnetic charges, analogous to the Coulomb law, the magnetic  force  is given by  $\vec {\cal F} (\vec r_n)=\fracc{1}{4\pi\mu_0}\fracc{m_p m_e}{r_n^2}\hat{r}_n$,\cite{griffithselectro2024} with $m_p$ and $m_e$ the magnetic charges of the permanent magnet and the electromagnet, respectively, and
$\hat{r}_n$ the unitary vector in the direction of  $\vec r_n$. Assuming small cantilever angles, $y_n\simeq d_0$, the horizontal force component per unit mass becomes
\begin{align}
F(u_n)=\fracc{m_p m_e}{4\pi\mu_0 M}\fracc{u_n}{(u_n^2+d_0^2)^{3/2}}\, .
\label{eq:horF}
\end{align}
The strength of the magnetic force is controlled by  the current $I_{EM}$ of the electromagnet. The first factor in Eq.\,(\ref{eq:horF}) can be identified with an interaction coefficient (force per unit mass),  $\chi (I_{EM})$,  which may be alternatively expressed as a function of the driving current $I_{EM}$ as  $\chi (I_{EM})=\chi_0+\chi_1  I_{EM}$. The constant term $\chi_0$ in the interaction coefficient is due to the ferromagnetic core of the electromagnet, while the variable term $\chi_1$ is due to the magnetic field created by the current of the electromagnet. Note that $\chi_0<0$, corresponding to an attractive force, and $\chi_1$ depends on the polarity of the current, which is such that in this work $\chi_1<0$ too.  The current consists of DC and AC (modulated) components, i.e.,
\begin{align}
&I_{EM}(n,t)=I_{DC}+I_{AC}\cos(hn-\Omega t)
\label{eq:current}
\end{align}
Therefore, the dynamical equations for an array of coupled cantilevers become:
\begin{align}
&\ddot u_n=-\omega_{0,0}^2 u_n+F(u_n)+C(u_{n+1}+u_{n-1}-2 u_n)
\label{eq:dynphys}
\end{align}
where $\omega_{0,0}$ is the natural frequency of an isolated cantilever (which, according to Euler--Bernouilli beam theory, depends on his geometry and the material properties), $C$ is a coupling constant, and the force is given by
\begin{align}
&F(u_n)=-\big(|\chi_0|+|\chi_1| I_{EM}(n,t)\big)\fracc{u_n}{(u_n^2+d_0^2)^{3/2}}\nonumber\\
&
\label{eq:dynphys2}
\end{align}
Typical values of the parameters, obtained in a previous experiment, are $\omega_{0,0}=2\pi\times 35.72$\,rad/s, $d_0=3$\,mm, $\chi_0=-4.71\times 10^{-5}$\,m$^3$s$^{-2}$, $\chi_1=-9.14\times 10^{-3}$m$^3$s$^{-2}$A$^{-1}$ and $C=284$\,s$^{-2}$.

Expanding the last term in Eq.(\ref{eq:dynphys2}), results in
\begin{align}
\fracc{u_n}{(u_n^2+d_0^2)^{3/2}}=\fracc{u_n}{d_0^3}-\fracc{3}{2}\fracc{u_n^3}{d_0^5}+o(u_n^5)\, ,
\end{align}
evidencing that the on-site potential is of soft type. Keeping only the linear term, we obtain the linearized equations:
\begin{align}
&\ddot u_n=-\omega_{0,0}^2 u_n\nonumber\\
&-\big(\fracc{|\chi_0|+|\chi_1| I_{DC}}{d_0^3} +\fracc{|\chi_1| I_{AC}}{d_0^3}\cos(hn-\Omega t)\big)u_n \nonumber\\
&+C(u_{n+1}+u_{n-1}-2 u_n)\, ,
\label{eq:dynphyslin}
\end{align}

The frequency of small oscillations of an isolated oscillator, in the absence of modulation, is  $\wp=(\omega_{0,0}^2+\fracc{|\chi_0|+|\chi_1| I_{DC}}{d_0^3})^{1/2}$.
We can re-scale the magnitudes in a relatively complicated way to obtain simpler equations,  by using $\tau=1/\wp$ and $\sigma$ such as $\sigma^3=\fracc{|\chi_0|+|\chi_1| I_{DC}}{\wp^2}$, as units of time and distance, respectively.  The scaled linear equations become

\begin{align}
\ddot u_n=-\wo^2 u_n -\delta\cos(h n-\Omega t) u_n+\kappa(u_{n+1} + u_{n-1}-2 u_n)\, ,
\label{eq:linearscaled}
\end{align}
with  $\wo=1$, that we keep for clarity, $\kappa=C/\wp^2$, and $\delta=\fracc{|\chi_1| I_{AC}}{\wp^2 d_0^3}$.

The non-linearized model, in the scaled variables, results
\begin{align}
&\ddot u_n=-\wo^2 u_n\nonumber\\
&-\big(-\fracc{1}{{\bar d}_0^3}u_n+(1+\delta{\bar d}_0^3\cos(h n-\Omega t))\fracc{u_n}{(u_n^2+{\bar d}_0^2)^{3/2}}\big)\nonumber\\
&+\kappa(u_{n+1} + u_{n-1}-2 u_n)\, ,
\label{eq:nonlinear1scaled}
\end{align}
for $\bar{d}_0=d_0/\sigma$.
The second term in parentheses is written in that way because it is completely nonlinear when expanding the terms in $u_n$. It also highlights that the linear frequency of the unmodulated system of the isolated oscillator is $\wo=1$.

The nonlinear equations in a more condensed form are: 
\begin{align}
&\ddot u_n=-\wo^2 u_n \nonumber\\
&-\big(-\delta_1 u_n+(1+\delta_2\cos(h n-\Omega t))\fracc{u_n}{(u_n^2+{\bar d}_0^2)^{3/2}}\big)\nonumber\\
&+\kappa(u_{n+1} + u_{n-1}-2 u_n)\,,
\label{eq:nonlinearscaled}
\end{align}
with $\delta_1=1/{\bar d}_0^3$ and  $\delta_2=\delta {\bar d}_0^3$.

For a representative set of parameters,  we fix DC and AC amplitudes at $I_{AC}=I_{DC}=12$\,mA, and in order to obtain that $\wo=1$, the scaling factors are $\tau=4.286$\,ms and $\sigma=1.423$\,mm, leading to  $\bar d_0=2.1087$, $\kappa=0.0051632$, $\delta_1=0.10665=$, $\delta_2=0.6996$.   


The linearization of \eqref{eq:nonlinearscaled} leads again to the system \eqref{eq:linearscaled}, with $\delta=\delta_2/{\bar d}_0^3=\delta_2\delta_1=0.07461$ for the chosen currents.

In the following we will write $d_0$ instead of $\bar d_0$ for simplicity.

The corresponding Hamiltonian is given by:\label{rev22-1B}
\begin{align}
&H=\sum_{n=1}^N e_n\quad \mathrm{with}\nonumber \\
&e_n=\fracc{1}{2}p_n^2+V_n(u_n,t)+\frac{1}{2}U(u_{n+1}-u_n)+\frac{1}{2}U(u_n-u_{n-1})\, ,
\end{align}
where $e_n$ is the local energy at site $n$, $V_n$ the site and time dependent on-site potential, and
$U(u_{n+1}-u_n)$ is the interaction energy between the variables $u_{n+1}$ and $u_n$.
We consider periodic boundary conditions so as $u_{N+1}=u_1$ and $u_0=u_N$.
The on-site potential $V_n$ is given by:
\begin{align}
V_n(u_n,t)=&\fracc{1}{2}\wo^2 u_n^2-\fracc{1}{2}\delta_1 u_n^2\nonumber \\
+&(1+\delta_2\cos(h n-\Omega t))\left(\frac{1}{d_0} -\frac{1}{\sqrt{u_n^2+d_0^2}}\right) \, .
\end{align}
The interaction potential is quadratic, given by:
\begin{align}
U(u_{n+1}-u_n)=\frac{1}{2}\kappa(u_{n+1}-u_n)^2 \, .
\end{align}
For small oscillations, the on-site potential becomes also quadratic:
\begin{align}
V_n^L(u_n,t)=\fracc{1}{2}\wo^2 u_n^2 +\fracc{1}{2}\delta\cos(hn-\Omega t))u_n^2 \, .
\end{align}
The Hamiltonian equations $\dot u_n=\fracc{H}{\partial p_n}=p_n$ and $\dot p_n=-\fracc{\partial H}{\partial u_n}$ are still valid and, therefore, the time derivative of the Hamiltonian becomes:
\begin{align}
 \fracc{\d H}{\d t}=\fracc{\partial H}{\partial t}=\delta_2\Omega\sum_n\sin(hn-\Omega t)\Big(\fracc{1}{d_0}-\fracc{1}{\sqrt{d_0^2+u_n^2}}\Big)\, .
 \label{eq:dHdt}
\end{align}

\begin{figure}[b]
\begin{center}
\includegraphics[width=\largefigure]{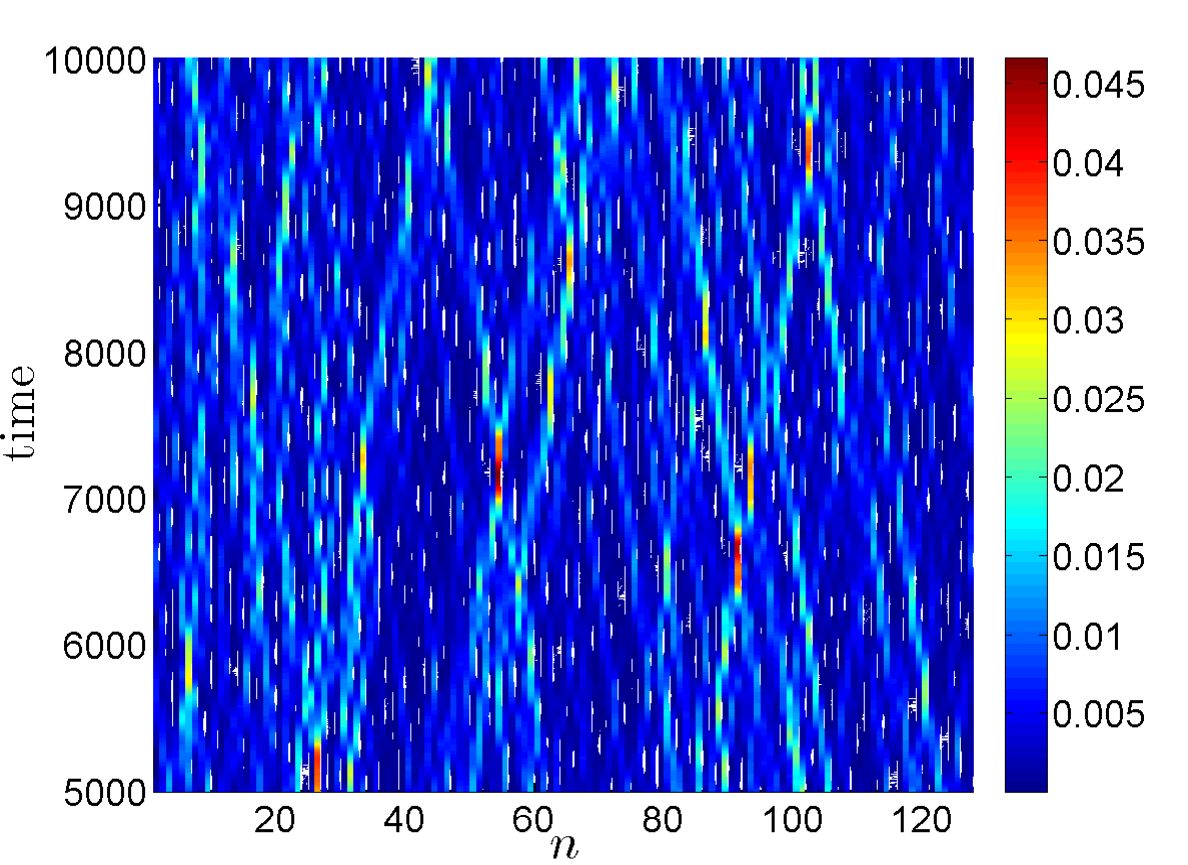} 
\includegraphics[width=\largefigure]{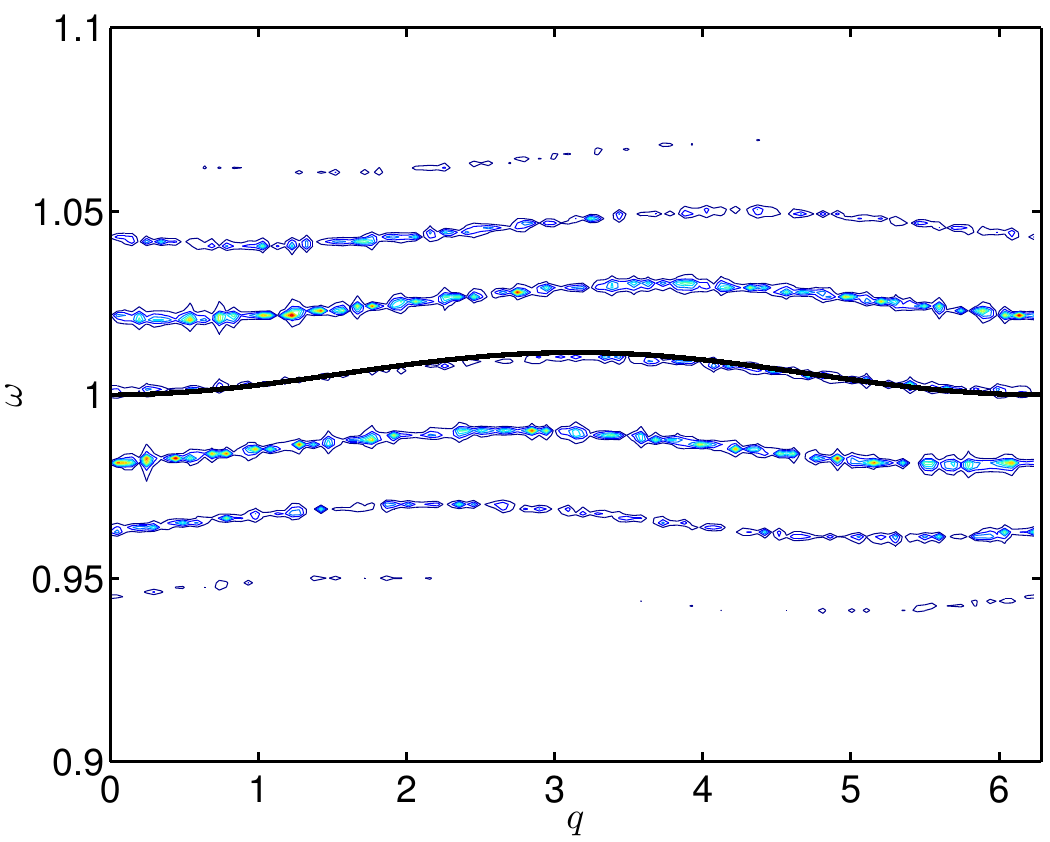} 
\end{center}
\caption{Numerical simulations after thermalization of the linear system. See
App.\,\ref{app:numericaldispersion_obtention} for details.
({ Top}) Energy density plot ({ Bottom})
Contour plot of the two-dimensional XTDFT of the coordinates $u_n(t)=u(n,t)$ as a function of $n$ and $t$.
Parameters: $\kappa=0.0052$, $\delta=0.075$, $\Omega=0.02$, $h=0.5$, $N=128$.
}
\label{fig_initial}
\end{figure}

In general, this derivative will not be zero and also it will not be periodic in time, implying that $H$ will not be periodic either. For $h=0$, the coefficients of $\sin(\Omega t)$ have all the same negative sign, and $\d H/\d t=0$ only if all $u_n=0$.
However, $H$ can be conserved at periodic times.  If $T_\text{m}=2\pi/\Omega$ is the modulation period and the coordinates $u_n$ are periodic with frequency $\wb$ and period $T_\text{b}=2\pi/\wb$, a sufficient condition for $H$ to be periodic with period $T_\text{H}$ is that there are integers $m_\text{m}$ and $m_\text{b}$ such that $T_\text{H}=m_\text{m}T_\text{m}=m_\text{b}T_\text{b}$ or, in terms of the frequencies:
\begin{align}
\fracc{\wb}{\Omega}=\fracc{m_\text{b}}{m_\text{m}}\, ,
\label{eq:commensurateH}
\end{align}
That is, the frequencies $\wb$ and $\Omega$ are commensurate. This is the case of
stationary breathers and other periodic solutions.

Note that for small oscillation Eq.\,\eqref{eq:dHdt} becomes:
\begin{align}
 \fracc{\d H}{\d t}=\fracc{\partial H}{\partial t}=\delta\Omega\sum_n\sin(hn-\Omega t)\frac{1}{2}u_n^2\,.
 \label{eq:dHdtlinear}
\end{align}

\section{Effect of space-time modulation on the phonon dispersion relation}
\label{sec:dispersionrelation}
In this section, we present the main consequences of space-time modulation of the linear system \eqref{eq:linearscaled}. They are also valid for small displacements $u_n(t)$ in the nonlinear model \eqref{eq:nonlinearscaled}.  Appendix\,\ref{app:thermal} explains the concept of thermalization and numerical experiments to observe the phonon dispersion relation (PDR). The relatively detailed mathematical derivations based on Bloch theorem are included in the App.\,\ref{app:dispersion}.

For $\delta=0$, that is, without modulation, the phonon dispersion relation is given by $\omega_q^2=\wo^2+2\kappa(1-\cos(q))$ as a function of the wave number $q$ of the phonon.
In most of this work, we use the physical value of the coupling parameter $\kappa=0.0052$, with the phonon frequencies between
$\omega_0=1$ and  $\sqrt{\wo^2+4\kappa}=1.010$. The exception is this section, where for some simulations we choose a larger value $\kappa=0.10$, with the phonon band between the frequencies 1 and 1.183, in order to have a good visualization of the shape of the original phonon band and the new bands that appear with the modulation. The parameters are specified in the captions.\label{rev22-4}

Numerically, the phonon spectrum can be obtained using the two-dimensional  discrete Fourier transform of $u_n(t)$ in space and time (XTDFT) as explained in App.\,\ref{app:numericaldispersion_obtention}.

For modulation frequencies $\Omega$ that are a few times smaller than $\wo$, the main effect is the emergence of several replicas of the phonon dispersion curves,
whose maxima are shifted and vertically displaced (in frequency) relative to the unmodulated dispersion curves, as shown in the bottom panel of Fig.\,\ref{fig_initial}.
The energy density plot for the same simulation is also presented in the upper panel of Fig.\,\ref{fig_initial}.

\begin{figure}[t]
\begin{center}
\includegraphics[width=\largefigure]{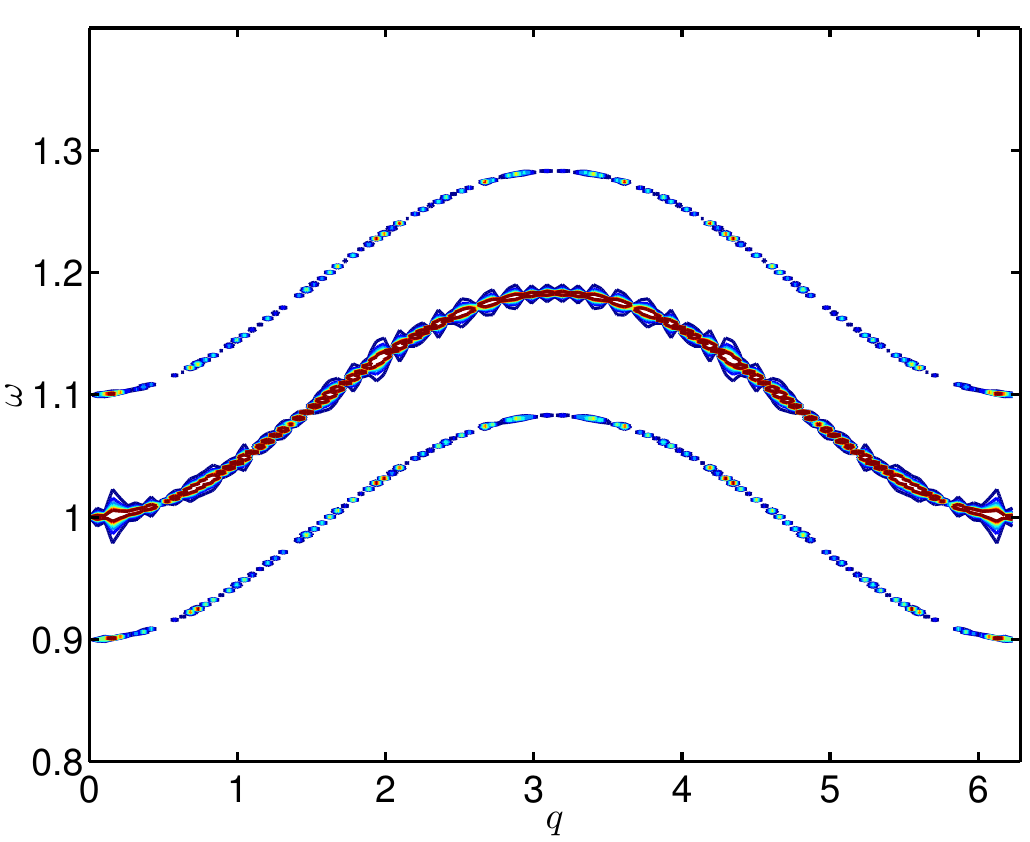} 
\includegraphics[width=\largefigure]{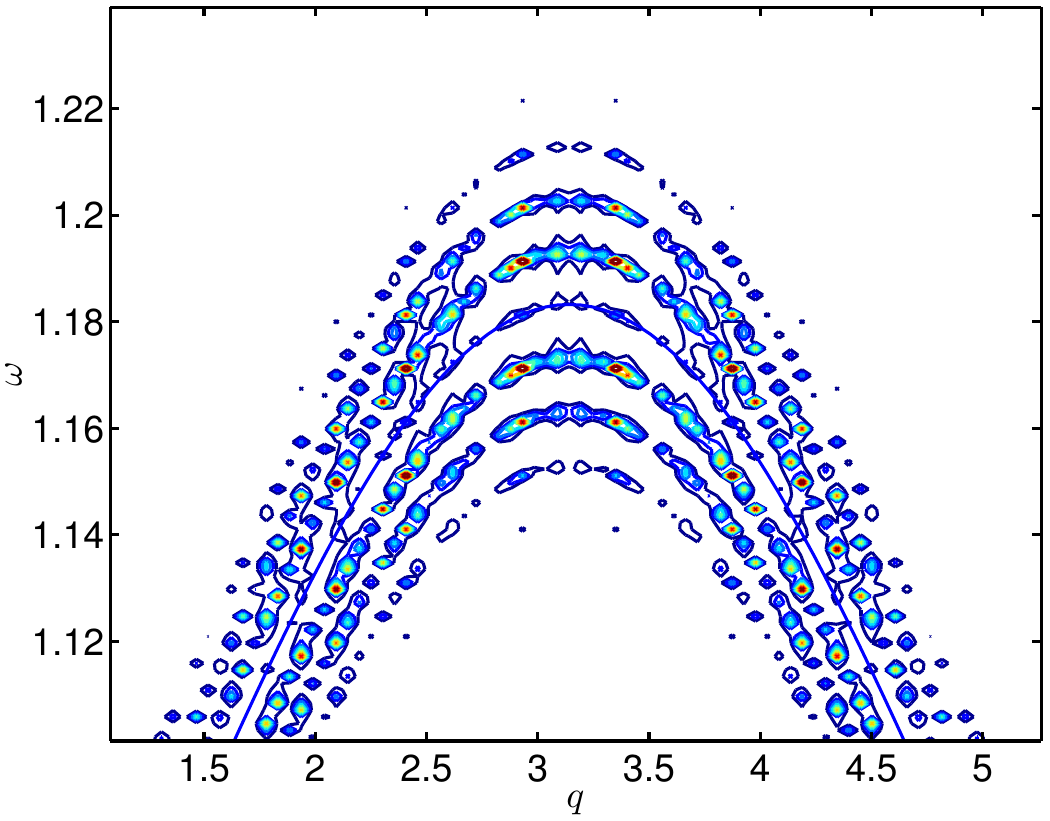} 
\end{center}
\caption{Contour plot of the XTDFT after thermalization of the linear system with only time modulation. See
 App.\,\ref{app:numericaldispersion_obtention} for details.
 Parameters: $\kappa=0.10$, $\delta=0.075$, $h=0$, N=120 (Top panel) $\Omega=0.1$; (Bottom panel) $\Omega=0.01$.
 } 
\label{fig_timemod}
\end{figure}

In App.\,\ref{app:timemodulation}, it is deduced that to first order approximation, the effect of time modulation alone is given by
\begin{align}
\omega= -m\Omega +\sqrt{\wo^2+2\kappa (1-\cos(q))}\, ,
\label{eq_phonon_timemod}
\end{align}
with $m=0,\pm1,\pm 2, \pm 3,\dots$.

Figure\,\ref{fig_timemod} presents two plots of the PDR obtained numerically for two different values of $\Omega$.

If there are both time and space modulation, it is also deduced in App.\,\ref{app:space-time} that the phonon bands are given by:
\begin{align}
\omega= -m\Omega +\sqrt{\wo^2+2\kappa (1-\cos(q+m h))}\, .
\label{eq:phonon_spmodulation}
\end{align}

\begin{figure}[t]
\begin{center}
\includegraphics[width=\largefigure]{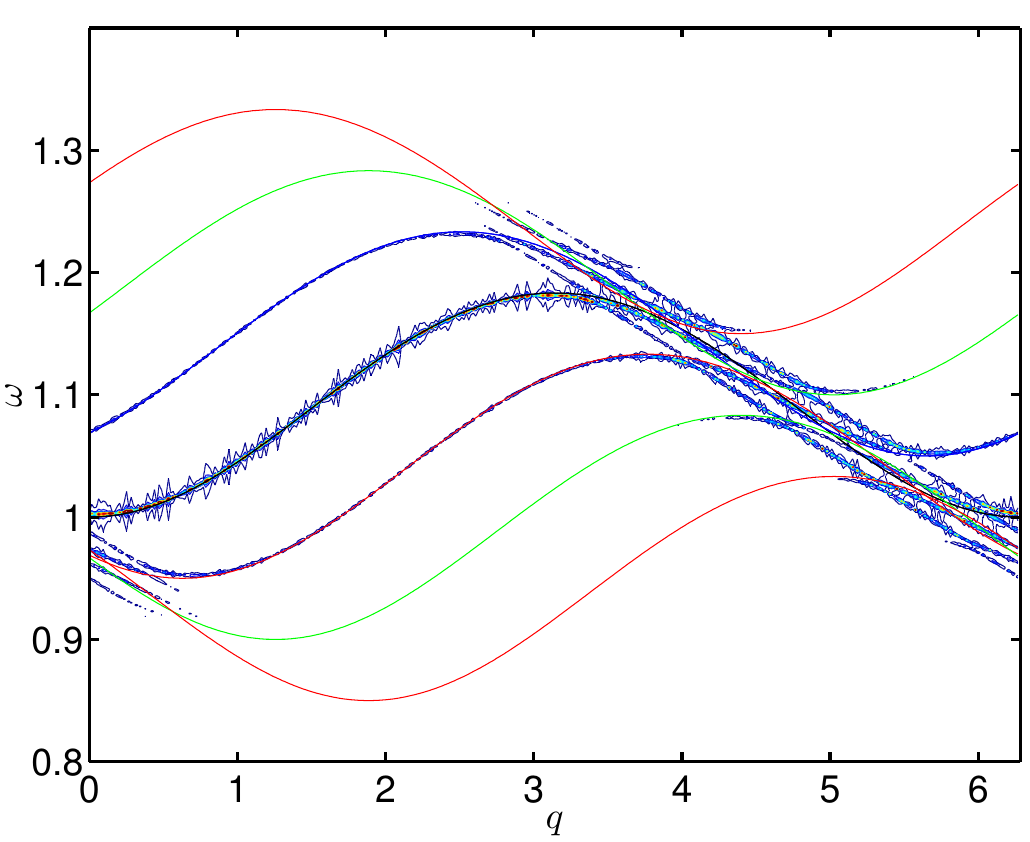}
\end{center}
\caption{Contour plot of the XTDFT after thermalization of the linear system with both time and space modulation, together with the theoretical dispersion curves. See App.\,\ref{app:numericaldispersion_obtention} and \ref{app:space-time} for details. Parameters: $\delta=0.05$, $\Omega=0.05$, $\kappa=0.10$, $h=0.5$, N=120. Note that there is an area to
the right where the harmonics are very close and become mixed.
 }
\label{fig_mod_spt}
\end{figure}

Figure\,\ref{fig_mod_spt} represents both the theoretical and numerically obtained PDR, showing a very good agreement.

The derivations presented in Appendix\,\ref{app:dispersion} cannot predict the intensity of the various PDR. In App.\,\ref{app:onlyspacededuction}, it is also deduced the effect of symmetry breaking of the space invariance by only space modulation.

\section{Breathers with space-time modulation}
\label{sec:theory_ILM}
In this section, we study the conditions for breather existence in a modulated system, first reviewing the theory of exact moving breathers and adapting it to space-time modulation and to only time modulation.

\subsection{Exact traveling breathers}
\label{sec:exactmoving}
The theory of moving breathers was developed by Aubry and coauthors\cite{chen-aubry-tsironis-PRE77-1996,aubry-cretegny-physd119-1998} and by
Flach and Kladko\,\cite{flach-kladko-physD127-1999}\label{rev22-5} among others. We will refer here to a formulation very similar to the latter, developed by some of the authors to interpret exact moving breathers in the $q-\omega$ space.\cite{archilla2019}  Other articles related to that formulation are Ref.\,\onlinecite{james-sire2005} with a rigorous mathematical treatment, and Ref.\,\onlinecite{gomez-gardenes2004} applied to the DNLS, where the concept of resonant lines (see below) appear. Here, we adapt the formulation in  Ref.\,\onlinecite{archilla2019} to modulated systems.

Let us suppose that we have a space-time modulated system with phase $\phi=hn-\Omega t$ and velocity $V_m=h/\Omega$. An exact breather will be a solution of the form:\cite{flach-kladko-physD127-1999}
\begin{align}
u(n,t)=f(n-V_b t,\omega_i t)\, ,
\label{eq:unt}
\end{align}
where the function $f$ is a localized function of its first argument and a $2\pi$ periodic function of the second,
and $V_b$ is the velocity of the breather. If the breather is exact, there exists a minimal time $T_F$, called the fundamental time, and an integer $s$, called the step, such after a time $T_F$, the breather reproduces itself displaced a distance $s$ in lattice units. Then, $V_b=s/T_F$, and the fundamental frequency is defined as
$\omega_F=2\pi/T_F$. The frequency $\omega_i$, is the moving frame frequency\,\cite{archilla2019} also called internal frequency.\cite{flach-kladko-physD127-1999}

The condition for $u(n,t)$ to be exact is that $u(n+s,t+T_F)=u(n,t)$, which implies that $s=V_b T_F$ and $\omega_i=m\omega_F$.
Harmonic functions with the same symmetry (same step and fundamental frequency) are called resonant harmonics and form a basis for the breather. They can be written as
\begin{align}
&\exp(\ii(q[n-V_bt])\exp(-\ii m\omega_F t)=\exp(\ii[ q n-\omega_L t]) \quad \nonumber \\
&\mathrm{with}\quad \omega_L=qV_b+m\omega_F\, .
\end{align}
where $\omega_L$ and $m\omega_F$  are the laboratory and moving frame frequencies, respectively, of the resonant harmonic.

The resonant harmonics form straight lines within the $(q,\omega)$ space, called {\em resonant lines}. Their slope is $V_b$ and their intercept at the axis $q=0$ is $m\omega_F$, i.e., the moving frame frequency. Therefore, all the harmonics in a resonant line have the same moving frame frequency $m\omega_F$, with the integer $m$ indexing the lines.

The breather frequencies are inside one of those lines called the {\em breather line} with $m=m_\text{b}$. The intercept of the breather line with the axis $q=0$ is the moving frame frequency of the breather  $\wb=m_\text{b}\omega_F$. All the harmonic waves in the same line have the same frequency in the moving frame and propagate with the same velocity $V_b$,  which explains the persistence of the breather.

For a soft potential as in our case, the breather line will be below the phonon band, and close to its minimum at $q=0$. If the breather line intersects the phonon band, the intersection phonon will be excited leading to a wing, an extended quasi-linear harmonic wave attached to the breather, with amplitude depending on the specific system and the frequency, and sometimes being zero.

\subsection{Application to space-time harmonically modulated systems}
\label{sec:application_space-time}
For a space-time harmonically modulated system with a term $\cos(hn-\Omega t)$. It is also necessary that the value of the modulating phase $\Phi=hn-\Omega t$ is also identical modulo $2\pi$ after the translation $s$ and time change $T_F$, because, if not, the forces will be different and the evolution of $u$ would be different.

Therefore:
\begin{align}
&h(n+s)-\Omega (t+T_F)=h n-\Omega t+2\pi r\,\Rightarrow \nonumber\\
&\Omega=\frac{h s-2\pi r}{T_F}\, \rightarrow \Omega=h V_b+m_\text{m}\omega_F\, ,
\end{align}
with $m_\text{m}$ an integer. This means that the modulating wave is a resonant harmonic with index $m_\text{m}$ and moving frame frequency $m_\text{m}\omega_F$ and, therefore, the breather and modulating waves have commensurate moving frame frequencies.

\noindent {\bf Lemma 1} {\em A necessary condition for an exact breather within a space-time modulated system with harmonic modulation is that the modulating harmonic is within a resonant line with the breather,  or, in other words, that their moving frame frequencies are both integer multiples of the fundamental frequency $\omega_F$. Therefore, $\omega_b$ and $\Omega$ are commensurate.}

Note, that an exact moving breather is not periodic in time and has not a definite frequency, this condition applies to the moving frame frequencies, and it is, therefore, different from Eq.\,\ref{eq:commensurateH}, which reappears below.

The immediate consequence for only time modulation is:

\noindent {\bf Lemma 2} For a harmonically, time-modulated system, a necessary condition for breather existence is that the
breather frequency $\omega_b$ and the modulating frequency $\Omega$ are commensurate: $\wb/\Omega=m_\text{b}/m_\text{m}$.

\begin{figure}[htb]
\begin{center}
\includegraphics[width=\largefigure]{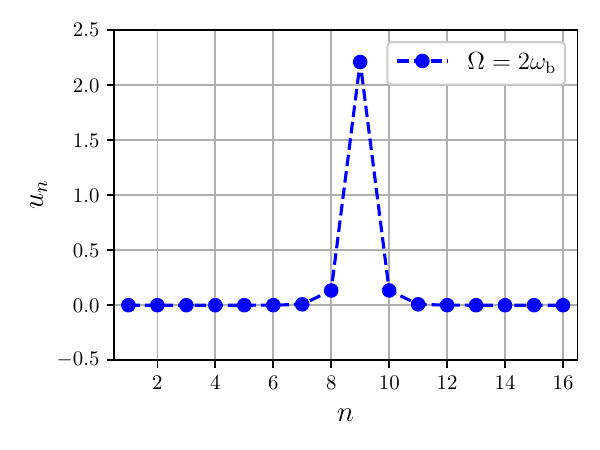}\\
\includegraphics[width=\largefigure]{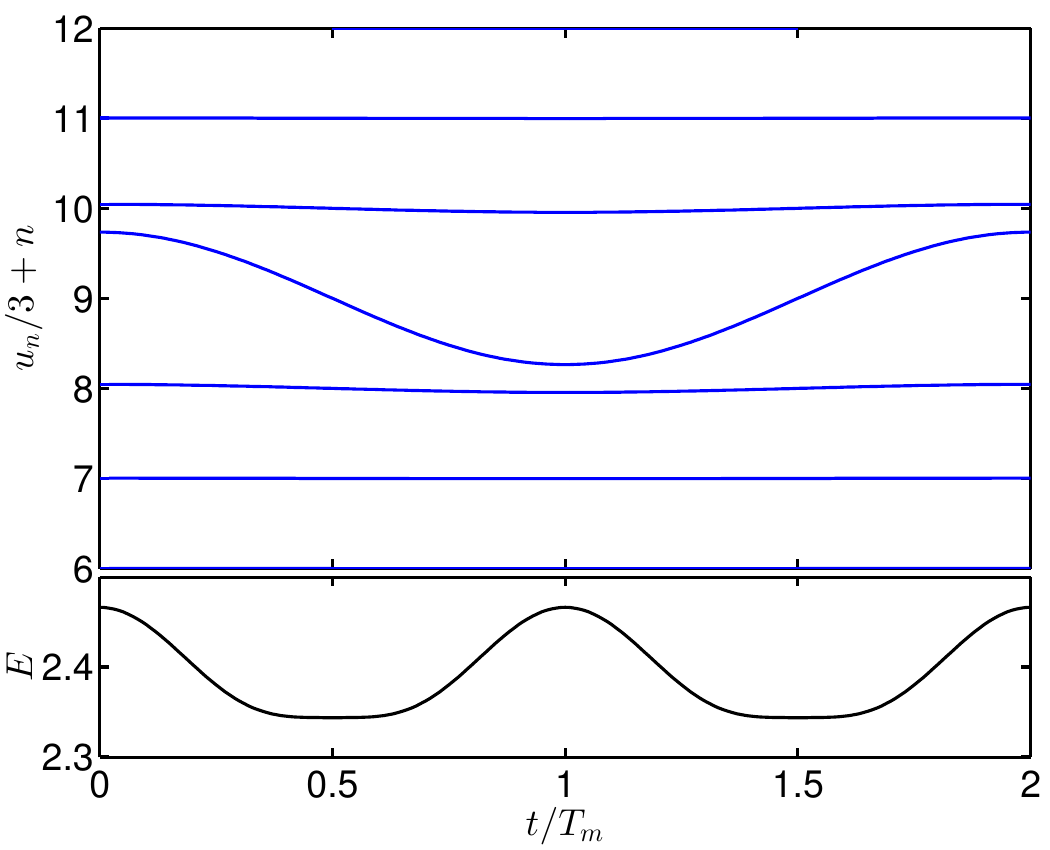}
\end{center}
\caption{ (Top) For $\Omega=2\wb$, profile of the stable breather. (Bottom) Coordinates and energy evolution in a breather period $T_\text{b}$.
Parameters: $\kappa=0.0052$, $\delta_1=0.1067$, $\delta_2=0.6996$, $\wb=0.98$.
}
\label{fig_o2wb_profile}
\end{figure}

\subsection{Breather obtention in the time-modulated system}
\label{sec:breatherobtention}
The most direct way is working in the real space with the Newton and shooting method. Testing some different localized initial conditions and observing an approximate periodical behavior with a period close to $T_\text{b}$, the intended one. The initial conditions are used as a seed for the Newton method, in order to obtain the initial conditions corresponding to the {\em exact} $T_\text{b}$-periodic behavior. {\em Exact} means with a precision of $10^{-15}$, that is, very close to the computer machine precision. The Newton method, a numerical application of the implicit function theorem, can be used because the system has no time invariance due to the time-periodic term in the potential, and a solution is unique in its neighborhood. The precision can be refined  further by working in the space of the frequencies. If we consider time-reversible solutions, they can be written as the inverse discrete Fourier transform $u_n(t)=\sum_{k=-k_\text{m}}^{k_\text{m}} z_{k,n} \exp(\ii k\wb t)$. Due to $u_n(t)$ being real and time-symmetric, $z_{k,n}^*=z_{-k,n}$ and $z_{k,n}=z_{-k,n}$ and real. Then,  $u_n(t)=z_{0,n}+\sum_{k=1}^{k_\text{m}} 2z_{k,n} \cos(k\wb t)$, and the Newton method can be coded to obtain the $k_\text{m}+1$ real coefficients $z_{k,n}$ for each site $n$.\label{rev22-6B} The value of $k_\text{m}=15$ provides excellent results.

The breather initial coordinates for the case $\Omega=2\omega_b$ are represented in Fig.\,\ref{fig_o2wb_profile}-top, the initial velocities being zero.

\subsection{Stability of breathers in the time-modulated system}
\label{sec:stability}

\begin{figure}[tbh]
\begin{center}
\includegraphics[width=\columnwidth]{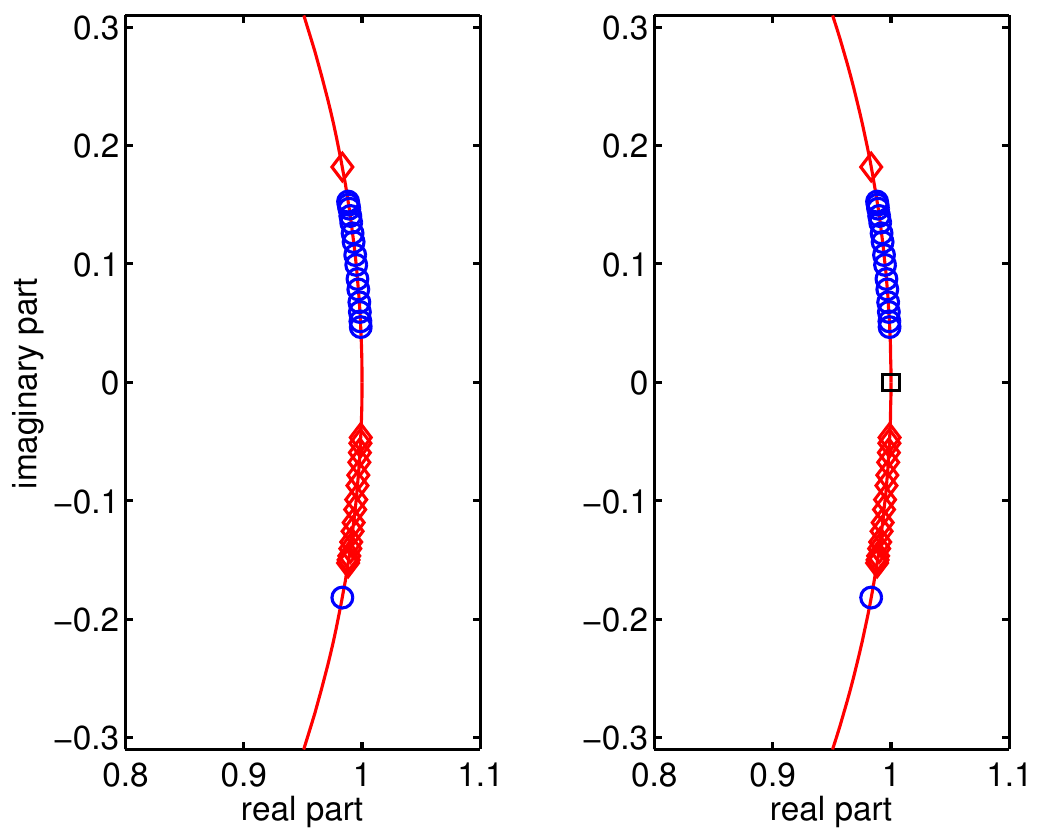}\\
\includegraphics[width=0.8\columnwidth]{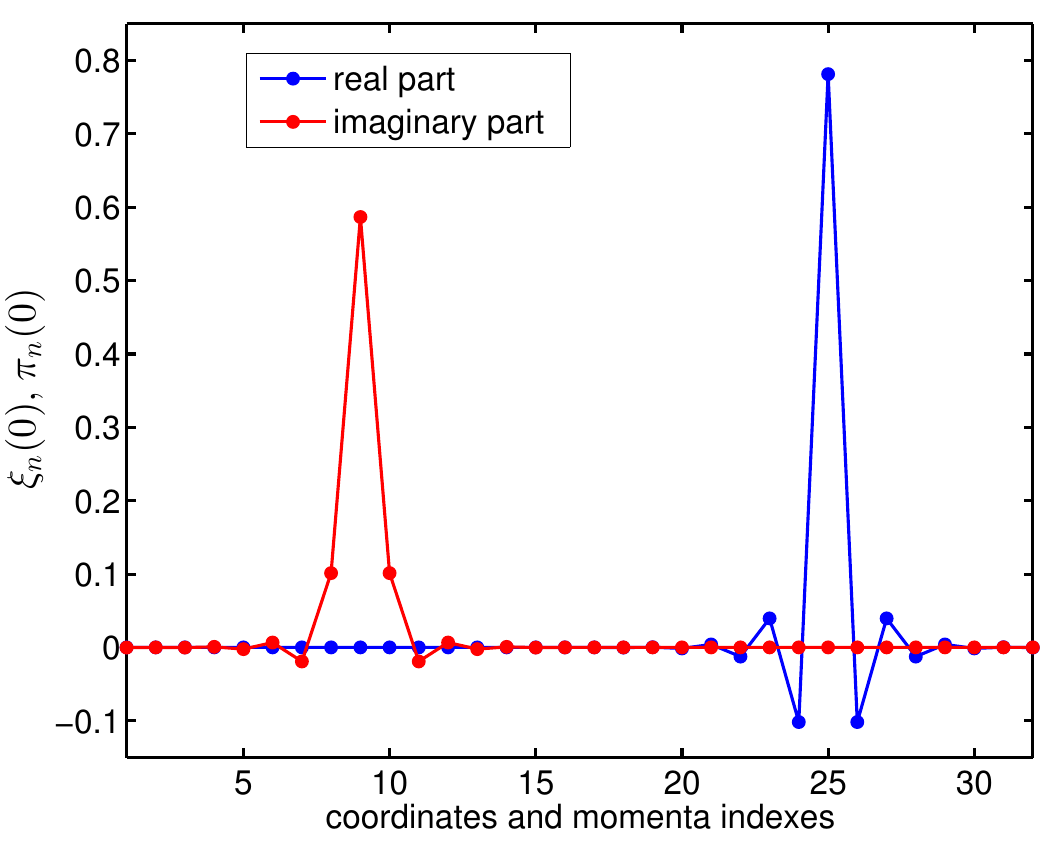}
\end{center}
\caption{ For $\Omega=2\wb$:(Top-left) Floquet multipliers for the non-autonomous system and (top-right) for the extended autonomous system.  Blue circles have positive Krein signature, red diamonds negative one, and black squares correspond to zero Krein signature. The multipliers are identical except for the double +1 of the extended system. (Bottom panel) Eigenvector corresponding to the isolated eigenvalues, related with the internal modes of the breather.
Parameters: $\kappa=0.0052$, $\delta_1=0.1067$, $\delta_2=0.6996$, $\wb=0.98$.
}
\label{fig_floquetextended}
\end{figure}

Let as suppose a breather solution 
 with  $T_\text{b}=n_T T_m$, with $n_T$ an integer larger than 1, the $T_m$-periodic system is also $T_\text{b}$ periodic, and to analyze the linear\label{rev22-3} stability we can construct the Floquet matrix integrating a small perturbation of the initial coordinates and momenta and observing the perturbation after $T=T_\text{b}$. For each changed coordinate or momentum we obtain a column of the Floquet matrix $F$. Its eigenvalues lie on the unit circle if the system is stable. For autonomous symplectic systems\,\cite{aubry2006} there are always two eigenvalues of $F$ at $+1$, corresponding to $u_t=\dot u$, the phase mode, as it represents the fact that if $u(t)$ is a $T_\text{b}$ periodic solution, $u(t+d t)$ is also a periodic solution, and the growth mode, indicating that around a solution there is another one with slightly different amplitude.

Those two eigenvalues do not generally appear for the non-autonomous system, as can be seen in Fig.\,\ref{fig_floquetextended}-top-left for $\Omega=2\wb$. The explanation is that if we write the dynamical equations \eqref{eq:nonlinearscaled} in a condensed form as:
\begin{align}
&\ddot u_n-L_n u_n
+\big(1+\delta_2\cos(\Omega t)\big)\fracc{\partial V(u_n)}{\partial u_n}=0\, ,
\label{eq:nonlinearcondensed}
\end{align}
where $L_n u= -\omega_1^2 u_n+\kappa (u_{n+1}+u_{n-1}-2u_n)$, with $\omega_1^2=\omega_0^2-\delta_1=\omega_0^2-1/d_0^3$, being the linear, unmodulated part of the dynamical system, with $V(0)=0$; and
$V(u_n)=1/d_0-1/\sqrt{d_0^2+u_n^2}$, the local non-modulated  on-site potential.

If $u_b(t)$ is a $T_\text{b}$-periodic solution of \eqref{eq:nonlinearcondensed}, and $u(t)=u_b(t)+\xi(t)$ a perturbed solution with $\xi(t)$ small, the evolution of $\xi$ is given by the Newton operator with zero eigenvalue, that is:
\begin{align}
N_n \xi&=\ddot \xi_n -L_n \xi \nonumber\\
&+\big(1+\delta_2\cos(\Omega t)\big)\fracc{\partial^2 V(u_{b,n}(t))}{\partial u_n^2}\xi_n=0\, .
\label{eq:newton}
\end{align}

For autonomous systems $\dot u_b$ is a solution of \eqref{eq:newton}, but not for non-autonomous ones, as the derivative of \eqref{eq:nonlinearcondensed} is given by:
\begin{align}
&\ddot{\dot{u}}_{b,n} -L_n \dot{u}_{b,n} \nonumber \\
&+\big(1+\delta_2\cos(\Omega t)\big)\fracc{\partial^2 V(u_{b,n}(t))}{\partial u_n^2}\dot u_{b,n}\nonumber \\
&-\Omega\delta_2\sin(\Omega t)\fracc{\partial V(u_{b,n}(t))}{\partial u_n}=0\, ,
\label{eq:ubderivative}
\end{align}
where the last term does not appear in \eqref{eq:newton}.

\label{rev22-10}
It can also be checked numerically that the phase mode $[\dot u_\text{b}(0),\ddot u_\text{b}(0)]^t=[0,\ddot u_\text{b}(0)]^t$ is not an eigenvector of the Floquet matrix $F$ with another eigenvalue.  The isolated eigenvalues outside the phonon band correspond to eigenvectors with the same localization of the breather, that are  internal modes of the breather. In particular, when they have the same symmetry of the breather, they are called breathing modes.\cite{Johansson2000} Simulations with the eigenvectors added as a perturbation to the breather show that they produce a beating of the amplitude of the breather which justify their name.

The double Floquet multiplier at +1 can be recovered within an autonomous extended system that includes the non-autonomous one, as shown below.
    For the different factors $n_T$, in $\Omega=n_T\wb$, we obtain different results that we will present later.

\subsection{Extended autonomous symplectic system}
The Hamiltonian structure of the dynamical equations, i.e.,  $\dot u_n=\partial H/\partial p_n$, $\dot p_n=-\partial H/\partial u_n$ is kept for the conjugate pairs of variables $u_n$, $p_n$, although the system is no longer autonomous. By defining $\qt=t$ or to be precise as  \eqref{eq:newton} is $T_\text{b}$-periodic, as $\qt=\mathrm{rem}(t,T_\text{b})$, the remainder of $t/T_\text{b}$, i.e.,  $\mathrm{rem}(nT_\text{b}+t,T_\text{b})=t$  and $\mathrm{rem}(-nT_\text{b}-t,T_\text{b})=-t$, both for $n>0$ and $t>0$, so as $\qt$ is $T_\text{b}$ periodic. Then $\dot \qt=1$ as usual, and the system with the extra variable has become autonomous, but not symplectic as $\qt$ has no conjugate variable yet. The extra variable $\pt$ to make the system symplectic is obtained by defining a new Hamiltonian $K=H+\pt$, with $\pt$ the conjugate momentum of $\qt$ as $\dot \qt=\partial K/\partial \pt=1$ as previously obtained. Then, $\dot \pt=-\partial K/\partial \qt=-\partial H/\partial \qt=-\partial H/\partial t=-\d H/\d t$, or $\pt=E_0-H$, with $E_0$ the initial energy at $t=0$.\cite{marthinsen2016} The extended system is now symplectic and $[\dot u_b,\dot \qt]$ is now solution of \eqref{eq:ubderivative} and being $T_\text{b}$ periodic, the Floquet multiplier $+1$ corresponding to the {\em phase mode} reappears. There is an extra $+1$ multiplier, because symplectiness implies that the multipliers are in pairs $\lambda, 1/\lambda$, the extra mode corresponding to a solution with slightly different energy or {\em growth mode}.

The extended system reveals a hidden symplectic structure of the autonomous system with the implication that linear\label{rev22-3} stability corresponds to all the Floquet multipliers being in the unit circle, but without the double multiplier of +1, which is recovered with the extended system. Figu\-re \ref{fig_floquetextended}-top-right shows these multipliers for $\Omega=2\wb$.

Note that these conclusions are valid both the cases $\Omega > wb$ and $\Omega \leq \wb$ studied in the next two sections taking due care of the time of integration. They are also valid for space-time modulated systems with trivial changes that we will detail elsewhere.

{\section{Breathers with time modulation and $\Omega=n_T \wb$}
\label{sec:timecrystals}
We consider the nonlinear dynamical system \eqref{eq:nonlinearscaled}  initially for the cases  $\wb=\Omega/n_T$ ($T\bre=n_T T_\text{m}$), and
$\wb=n_T\Omega$ ($T\bre=T_\text{m}/n_T$), $n_T$ an integer. The first one corresponds to the well known phenomenon of period doubling, tripling, etc., while the latter correspond to a breather that oscillates several times faster than the modulation.   According to the deduction in Sec.~\ref{sec:theory_ILM}, fractional frequencies are also possible, and we will consider them in Sec.\,\ref{sec:frequencydependence} for different breather frequencies.\label{rev22-11}

There is no dissipation of energy, but as the system is parametrically forced, the energy is conserved only at discrete times separated by the larger of the two periods, the breather period  $T_\text{b}$ and the modulating one $T_m$. This is shown in Fig.\,\ref{fig_o2wb_profile} for $T_\text{b}=2T_m$.

In this section, we consider  $\wb=\Omega/n_T$   and leave the condition of $\wb>\Omega$ for the next section.

The dynamical equations\,\eqref{eq:nonlinearscaled} are invariant under the transformation $t \rightarrow t+T_m$, therefore, if a phonon $(q,\omega)$, i.e., $u=\exp(\ii[qn-\omega t])$ is a solution of the linearized equation, then also $\exp(\ii  m\Omega t) u$ is also a solution, or, in other words, there is another phonon band with frequencies $\omega'=\omega+m\Omega$, with $m$ a positive or negative integer. But there are also two phonon bands, with positive and negative frequencies. Then, if there is a phonon $(q,\omega)$, there also exists the phonon $(q,\Omega-\omega)$. As the unmodulated phonon band for $\kappa=0.052$ is in  $[\wo=1,\omega_\text{max}=1.01]$, for $\Omega\simeq 2$, a new phonon band appears in
$[\Omega-\omega_\text{max}\simeq 0.98,\Omega-\wo\simeq 1]$. The phonon bands in the modulated system change position, and $\Omega=2\wb=1.96$, but those bands are very close with the possibility of interfering, depending on the exact value of $\wb$ and actual modulated phonons. This problem does not exist for $\Omega \geq 3$ as the band $\Omega-\omega$, for $\omega$ in the unmodulated phonon band is much more separated. \label{rev22-7}

For $\Omega=2\wb$, and changing slightly $\wb$,  there is an interval where the two phonon  bands coincide and then the system becomes stimulated by $\Omega$ and the variables diverge, as it is also shown in Sec.\,\ref{sec:frequencydependence}. A dissipation term can be added, but
as this is a very special case, we prefer not to modify the system and simply avoid those combinationsof frequencies.

\subsection{Period doubling $\Omega=2 \wb$}
\label{sec:period-doubling}
The subject of time-modulation is of recent interest in photonic topological insulators. For example, Floquet solitons, for which the intensity repeats after each driving period, were found experimentally in optical waveguide arrays with Kerr nonlinearity\,\cite{mukherjee-science2020} and period-doubled Floquet solitons were observed in a photonic topological insulator formed by an array of waveguides with a honeycomb pattern.\cite{mukherjee-rechtsman2023} For these solutions, the intensity period is doubled. The same happens in a helically modulated honeycomb lattice as well and a sinusoidally driven SSH lattice. 
\label{rev22-8} 
These references show that the period-doubling phenomenon can appear in many different physical systems of interest and, in particular, photonic systems are ideal experimental devices to observe the connection between time modulation, topology, and nonlinearity.

Period doubling also appears in our system. In Fig.\,\ref{fig_o2wb_profile}-bottom we can see the evolution of the breather center and its neighboring sites, also the evolution of the energy, which is $T_m$-periodic and, therefore, also $T_\text{b}$-periodic. This is a consequence of $u_n(T_\text{b}/2)=-u_n(0)$, $p_n(T_\text{b}/2)=-p_n(0)=0$,  the energy being a function of $u_n^2$ and $\cos(\Omega t)=1$ at values of $t$ that are integer multiples of $T_{b}/2$.
There is no net gain or loss of energy every $T_\text{b}/2$ with a relative precision of $10^{-9}$ after 100 breather periods using a symplectic integrator.\cite{sanzsernabook1994} \label{rev22-9}

The Floquet eigenvalues are also pictured in Fig.\,\ref{fig_floquetextended}-top-left. The eigenvalues corresponding to the phonons form a pair of complex conjugate bands width frequencies $\omega_n=\wb+\theta_n/T_\text{b}$, as they have the same Krein signature,\cite{aubry2006} they form a structurally stable subset. There is also a close but separate eigenvalue with a localized eigenvector shown in Fig.\,\ref{fig_floquetextended}-bottom, \label{rev22-10}that corresponds to the breathing mode, a type of internal breather mode as explained in Sec.\,\ref{sec:stability} . It has the opposite Krein signature of the phonons, meaning it can collide with the phonon with the closest frequency leaving the unit circle, which indicates that the breather is unstable and another solution exists. As shown in Sect.\,\ref{sec:frequencydependence}, in some cases the new solution corresponds to a breather  in quadrature, i.e., with a difference of phase of $\pi/2$, with the modulation function.\label{rev22-11}

\begin{figure}[b]
\begin{center}
\includegraphics[width=0.8\columnwidth]{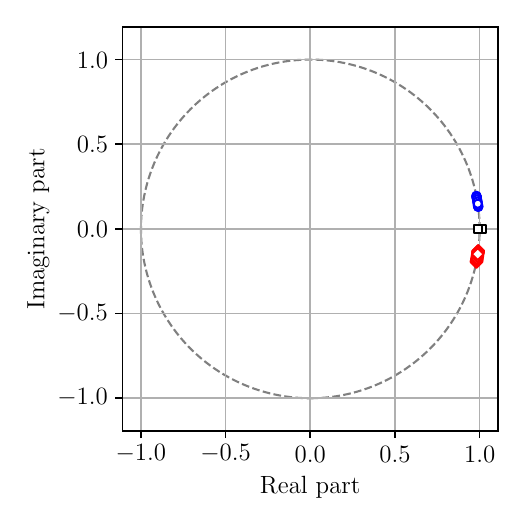}\\
\includegraphics[width=0.8\columnwidth]{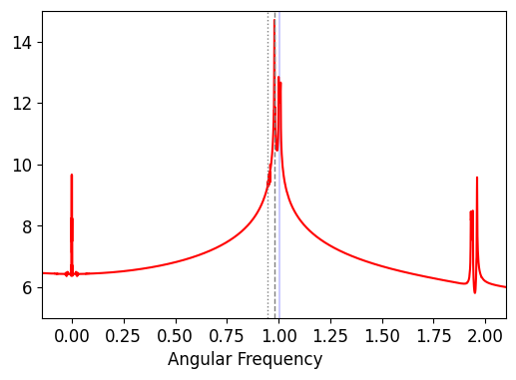}
\end{center}
\caption{(Top) Floquet eigenvalues for a breather with $\Omega=3\wb$.
(Bottom) Fourier spectra of the same breather embedded in a noisy system with $N=160$ cantilevers.
Parameters: $\kappa=0.0052$, $\delta_1=0.1067$, $\delta_2=0.6996$, $\wb=0.98$.
}
\label{fig_wb3Ofloq_spec}
\end{figure}

\subsection{Cases $\wb=3\Omega$ to $\wb=10\Omega$}

For $\wb=3\Omega$, the breather is slightly unstable with two real multipliers very close to +1. In spite of that, simulations can be done to simulate the breather evolution for a relatively long time.  The Floquet multipliers are represented in Fig.\,\ref{fig_wb3Ofloq_spec} and the Fourier spectrum in a system with noise and $N=160$ oscillators is also shown.

For $\wb=n_T\Omega$, width $n_T=1\dots 10$, we have constructed breathers, for $n_T=3,4,8$, breathers are unstable, while the rest are stable, therefore, there is no apparent pattern. For example, for $n_T=3$, there are  two real eigenvalues slightly separated from $+1$, but, in spite of that, the breather has a long life.
In Sec.\,\ref{sec:frequencydependence}, there is also more information on the stability of this type of breathers  and its evolution with the frequency.\label{ref22-11}

\section{Breathers with  $\Omega=\wb/n_T$}
\label{sec:antitimecrystals}

A peculiarity of this case is that the linear\label{rev22-3} stability of the breather and the Floquet matrix has to be calculated for a time equal to the modulation period $T_m=n_T T\bre$ as the coordinates, and velocities have to repeat, but also the forces which depend on time and
only repeat after $T_m$.
Therefore, the phonon band appears with Floquet exponents multiplied by $n_T$ and spread out as they are given by $\theta_q=\omega_q T_m =n_T\omega_q T_\text{b}$ for a phonon of frequency $\omega_q$.

\begin{figure}[thb]
\begin{center}
\includegraphics[width=\largefigure]{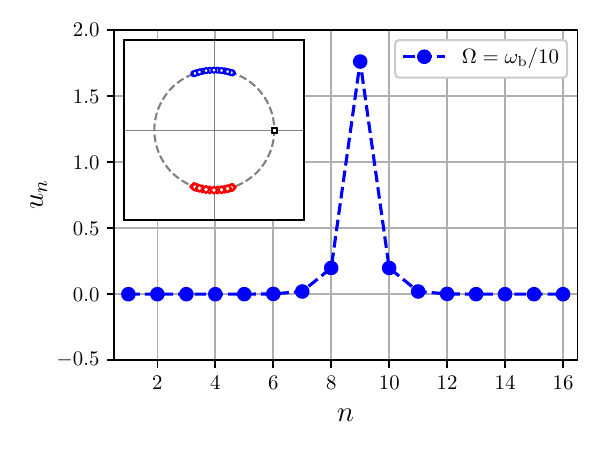} 
\includegraphics[width=\largefigure]{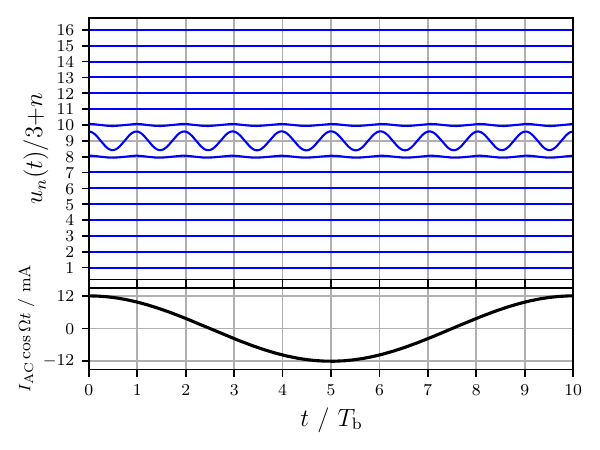}
\end{center}
\caption{(Top) Profile and Floquet eigenvalues for a breather with $\Omega=\wb/10$. The Krein signature of the Floquet eigenvalues is coded blue(+), red(-), black (zero).
(Bottom) Evolution of the coordinates and the AC current.
Parameters: $\kappa=0.0052$, $\delta_1=0.1067$, $\delta_2=0.6996$, $\wb=0.98$.
}
\label{fig_O_wdiv10_profile_floq}
\end{figure}

In this section, we separate the study of the Sievers-Takeno mode\,\cite{sieverstakeno1988} or single breather, from the Page mode\,\cite{page1990} or double breather.
Values $n_T=1,2,\dots,32$ are considered and breathers obtained. For $n_T=32,\dots,40$, phonobreathers\,\cite{marinaubry1996} appear as explained in Sec.\,\ref{sec:phonobreather}.
\begin{figure}[htb]
\begin{center}
\includegraphics[width=\halffigure]{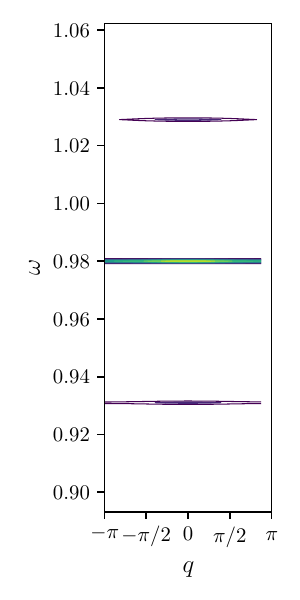}
\includegraphics[width=\halffigure]{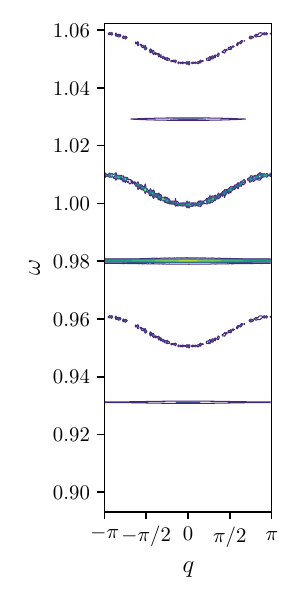}\\
\includegraphics[width=\largefigure,trim={0.2cm 0 0.2cm 0},clip]{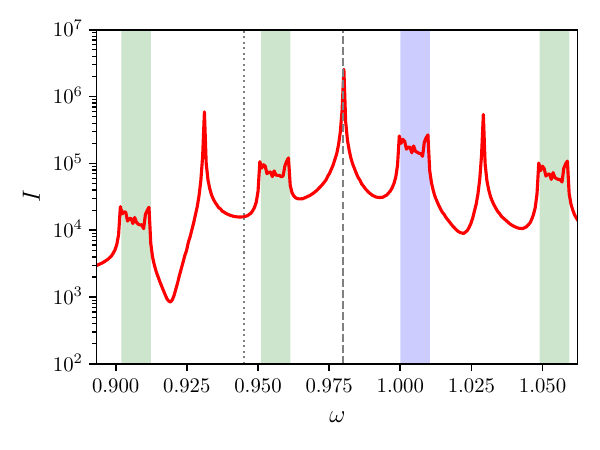}
\end{center}
\caption{ Case  $\Omega=\wb/20$.  (Top)~Frequency-momenta representation with scaled frequency. To the left for $N=16$ sites, and to the right for $N=160$ and noise added. See text.
(Bottom)~Spectra of the Fourier intensities in arbitrary units with respect to the scaled frequency. See text.
Parameters: $\kappa=0.0052$, $\delta_1=0.1067$, $\delta_2=0.6996$, $\wb=0.98$.
}
\label{fig_O_wdiv20_fft}
\end{figure}

\subsection{Sievers-Takeno mode: single ILM with  $\wb=n_T\Omega$}
\label{sec:sievers-takeno}
A stationary, single ILM or breather is a localized periodic vibration with a given frequency $\wb$ and with a single site, that we call the central one, having a large amplitude with respect to the phonons, and the amplitude of the neighbors diminishing rapidly with their distance to the center. They are often constructed from the anticontinuous limit with a single excited oscillator and the others at rest.\cite{mackayaubry94,marinaubry1996,flachgorbach2008}

In our
system, the unmodulated system dispersion relation is given by $\omega^2=\wo^2+2\kappa(1-\cos(q))$, with a minimum frequency $\wo=1$.  As the nonlinear part of the on-site potential is soft, because the dominant term is $o(3)$, stationary breathers or ILMs will have a frequency $\wb$ below the minimum frequency $\wo$. As demonstrated in Sec.\,\ref{sec:theory_ILM}, for a time-modulated system, the modulating frequency $\Omega$, is commensurate with $\wb$, and in this subsection, we consider the case $\Omega=\wb/n_T$. By using the shooting method, we tried and obtained exact ILMs for many values of $n_T$, as 1, 2, \dots 40. We present here only some results with $\wb=0.98$, that is, very close to the minimum frequency $\wo=1$ for the unmodulated system.

We construct the ILM in a small lattice with $N=16$ cantilevers, and check its linear\label{rev22-3} stability by constructing the monodromy matrix and obtaining the Floquet multipliers and exponents.\cite{aubry2006} Then, we embed the initial positions within a larger lattice with $N=160$ particles with some small random velocities, let it evolve and perform the XTDFT to check the results. An example $\Omega=\wb/10$ with $\wb=0.98$ can be seen in Fig.\,\ref{fig_O_wdiv10_profile_floq}. The top panel shows the profile at $t=0$ and the Floquet eigenvalues, showing its stability. The bottom panel illustrates the evolution of the variables for a period of the AC current.

We choose the case $\Omega=\wb/20$ with $\wb=0.98$, to illustrate the frequency-momenta representation as it is more easily seen.
The top-left panel of Fig.\,\ref{fig_O_wdiv20_fft}\label{rev22-13} shows the numerical XTDFT
(see App.\,\ref{app:numericaldispersion_obtention}) of an exact breather at zero temperature in a small system with $N=16$.  The breather appears as a horizontal line corresponding to the breather frequency $\wb=0.98$, with two other breather bands at frequencies $\wb\pm\Omega$. The lines are extended to almost all the momenta because the breather is extremely localized. No phonons appear.

\begin{figure}[t]
\begin{center}
\includegraphics[width=0.8\columnwidth]{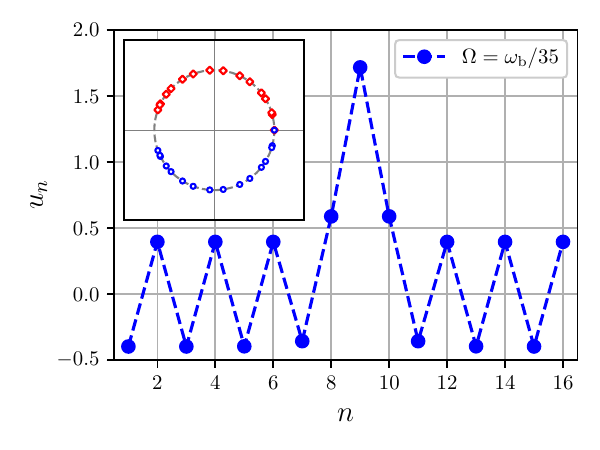}\\
\includegraphics[width=0.8\columnwidth]{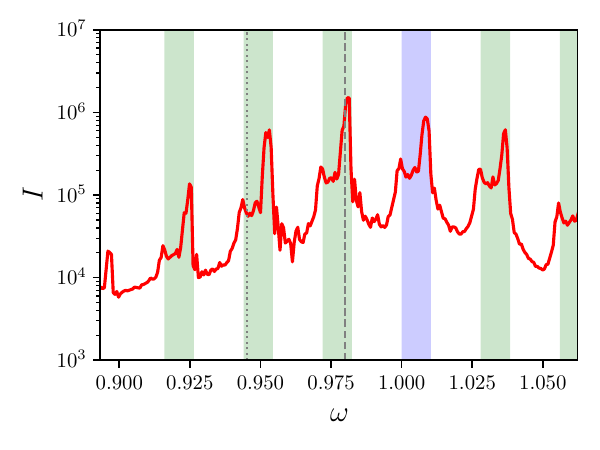}
\end{center}
\caption{ (Top) For $\Omega<\wb/33$, a phonobreather appears with a background wave to the ILM. The inset shows the Floquet eigenvalues with the Krein signature coded blue(+), red(-), black (zero). (Bottom) Fourier spectra. See text.
Parameters: $\kappa=0.0052$, $\delta_1=0.1067$, $\delta_2=0.6996$, $\wb=0.98$.
}
\label{fig_phonobreather}
\end{figure}

Figure\,\ref{fig_O_wdiv20_fft}-top-right shows the XTDFT of the same breather but located within a larger system with $N=160$ and thermalized with initial random velocities. The breather survives, and the phonon band becomes apparent. Two replicas of the phonon band can be seen with frequencies
$\omega=\pm \Omega + \sqrt{\wo^2+2\kappa (1-\cos(q)}$, and, also, tow replicas of the breather band with frequencies $\wb \pm \Omega$. Each breather-line replica is located below the corresponding phonon band replica. Other replicas with weaker intensity are out of the frame.

Figure\,\ref{fig_O_wdiv20_fft}-bottom shows
the numerical Fourier spectra obtained for the larger lattice with noise, adding up the intensities for all the particles at the same time.
We can also see that the
frequencies $\wb\pm \Omega$ appear displaced $\wo-\wb=1-0.98=0.02$ above and below the bottom of the secondary dispersion curves.

\begin{figure}[b]
\begin{center}
\includegraphics[width=0.8\columnwidth]{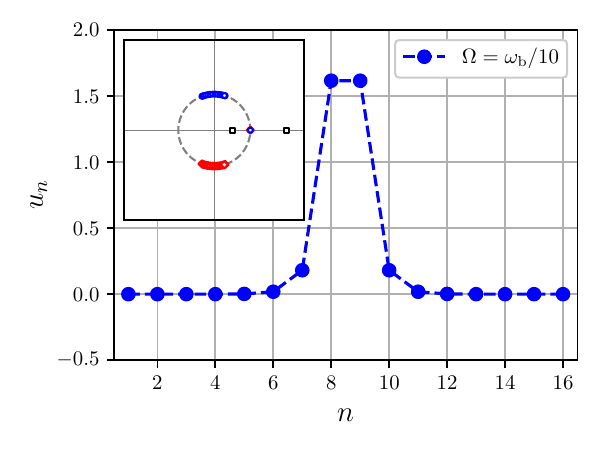}\\
\includegraphics[width=0.8\columnwidth]{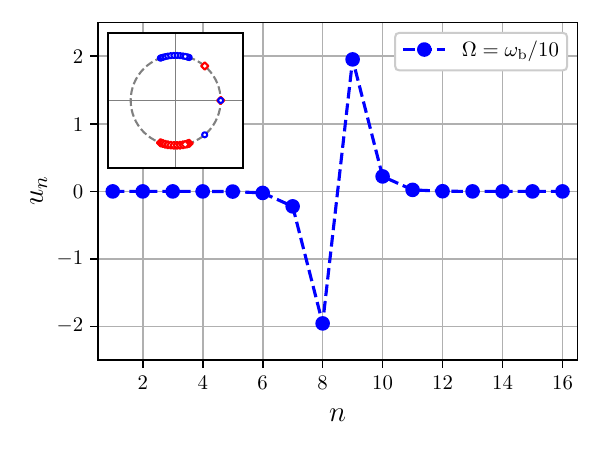}
\end{center}
\caption{ (Top) For $\Omega<\wb/10$, profile of the unstable in-phase Page mode. (Bottom) The stable anti-phase Page mode with the same parameters and modulating frequency.
The insets show the Floquet eigenvalues with the Krein signature coded blue(+), red(-), black (zero).
Parameters: $\kappa=0.0052$, $\delta_1=0.1067$, $\delta_2=0.6996$, $\wb=0.98$.
}
\label{fig_bothPage_prof_floq}
\end{figure}

\subsection{Phonobreathers: single ILM with phonon background for $\wb=n_T\Omega$}
\label{sec:phonobreather}
Although obtaining an ILM
for increasing $n_T$ is possible, as $\Omega$ becomes smaller, the upper secondary ILM band $\wb+\Omega$ collides with the upper frequency of the phonon band $\wmax=\sqrt{\wo^2+2\kappa(1-\cos(\pi))}=\sqrt{\wo^2+4\kappa}=\wb+\Omega$, or $\Omega=\wmax-\wb$ and $\wb/\Omega\simeq 32.3$, for the values  $\wo=1$ and $\kappa=0.0052$.  Then for $n_T>30$ the upper phonons are excited and a extended wave with $q=\pi$ profile appears as the background of the ILM, with smaller but comparable amplitude, forming a {\em phonobreather}.  A phonobreather is a periodic solution on a nonlinear system formed by a breather with a background of homogeneous amplitude. They were first described by Mar\'in and Aubry.\cite{marinaubry1996,aubry1997,marin-aubry-physicaD119-1998}
The background is not a phonon as it is not a solution of the linearized system and has a significant amplitude. It can also be described as a nonlinear extended wave.

The profile of the phonobreather  can be seen in Fig.\,\ref{fig_phonobreather}-top, and at the bottom of the same figure, the Fourier spectra, where the colliding of the top of the phonon bands can be seen merging with the ILM bands of upper order, while in Fig.\,\ref{fig_O_wdiv20_fft}-bottom they are well separated.

\subsection{Page modes with  $\wb=n_T\Omega$}
\label{sec:pagemode}
A Page mode\,\cite{page1990,flachgorbach2008} is a double breather or ILM with two sites vibrating with the same amplitude. It can be bond-symmetric if both sites oscillate in-phase (iP), or bond-antisymmetric if they oscillate in anti-phase (aP). The iP mode is unstable and the aP is stable for an unmodulated Klein Gordon system with soft potential.\cite{marinaubry-floria-physicaD113-1998}\label{rev22-14} Both two modes appear also in the modulated system with the same stability properties.
Both mode profiles and Floquet multipliers are represented in Fig.\,\ref{fig_bothPage_prof_floq} for $\wb= 10\Omega$. The XTDFT of the in-phase mode shows the appearance of another breather band closer to the bottom of the phonon band, which corresponds to the unstable mode. Similar phenomena to the single ILM occur, i.e., formation of replicas of the phonon band and ILM band displaced $\pm\Omega$ in the frequency space and the excitation of the background forming a phonobreather when the secondary ILM band interacts with another replica of the phonon band just below.

\begin{figure*}[p]
\begin{center}
\includegraphics[width=0.8\textwidth,clip]{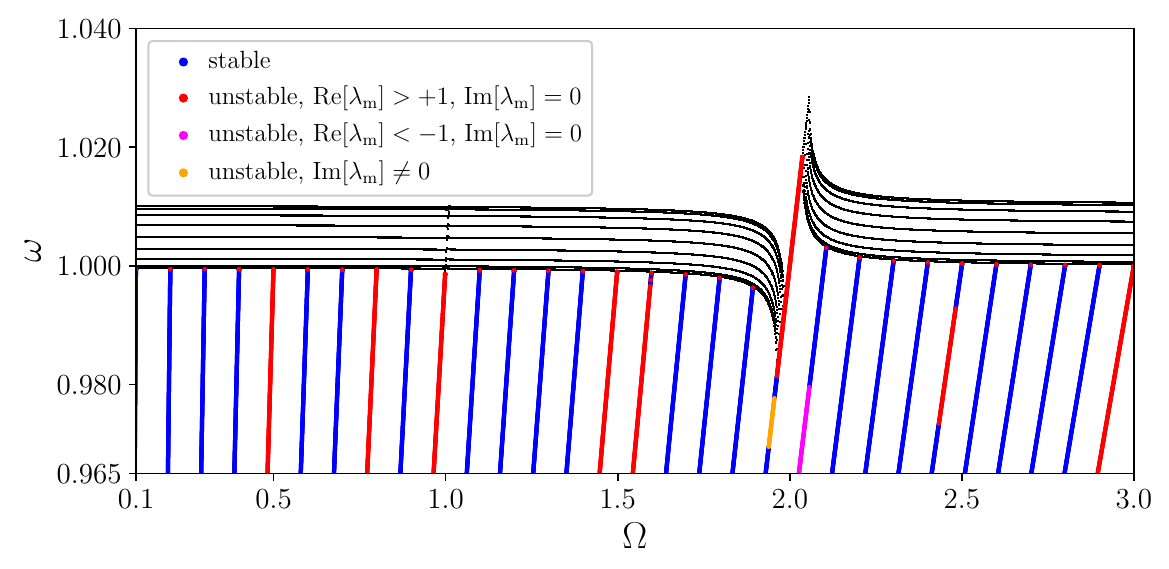}
\includegraphics[width=0.8\textwidth,clip]{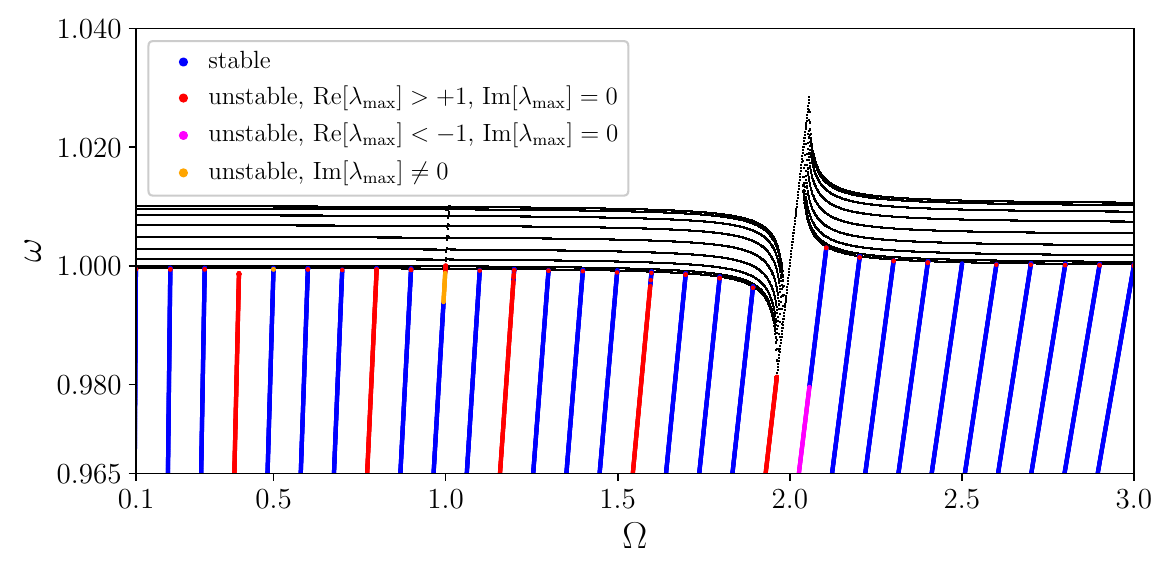}
\end{center}
\caption{(Top) ST in-phase breathers (Bottom) ST in-quadrature breathers.
Evolution of the frequencies of phonons and breathers with indication of their stability. $\lambda_{\rm max}$ denotes the eigenvalue with the maximum absolute value.
The breather/modulation frequency ratios correspond to $\wb/\Omega=m_\text{b}/m_\text{m}$ for  $m_\text{b}=10$ and $m_\text{m}=[2,\dots,30]$, that is,  $\wb/\Omega=m_\text{b}/m_\text{m}=[5, 3.333,\dots, 1, \dots, 0.5, \dots, 0.3333]$, corresponding to the slopes of the straight lines from left to right.\label{rev22-15}   Particular cases of interest are Floquet breathers with $\wb=\Omega$ and period-doubled breathers $\wb=\Omega/2$. See text for details.
Parameters: $\kappa=0.0052$, $\delta_1=0.1067$, $\delta_2=0.6996$.
}
\label{fig_phonon_band_breathers}
\end{figure*}

\begin{figure*}[p]
\begin{center}
\includegraphics[width=0.47\textwidth, clip, trim=0 8 0 40]{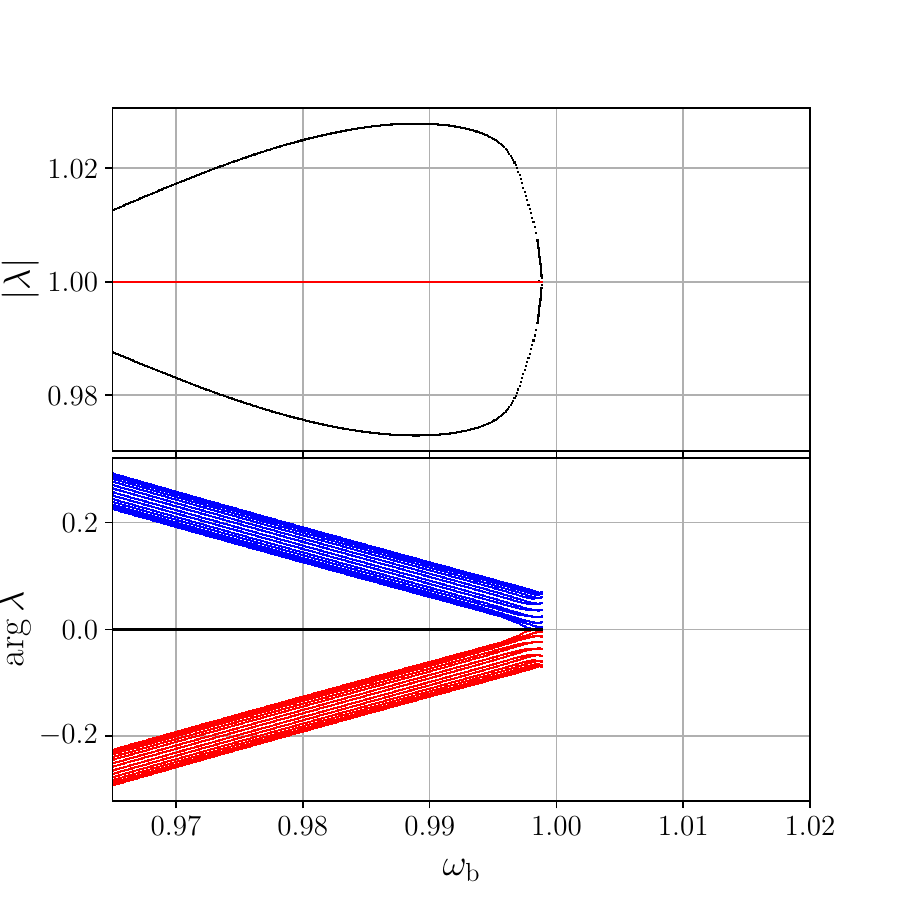}\mbox{}
\includegraphics[width=0.47\textwidth, clip, trim=0 8 0 40]{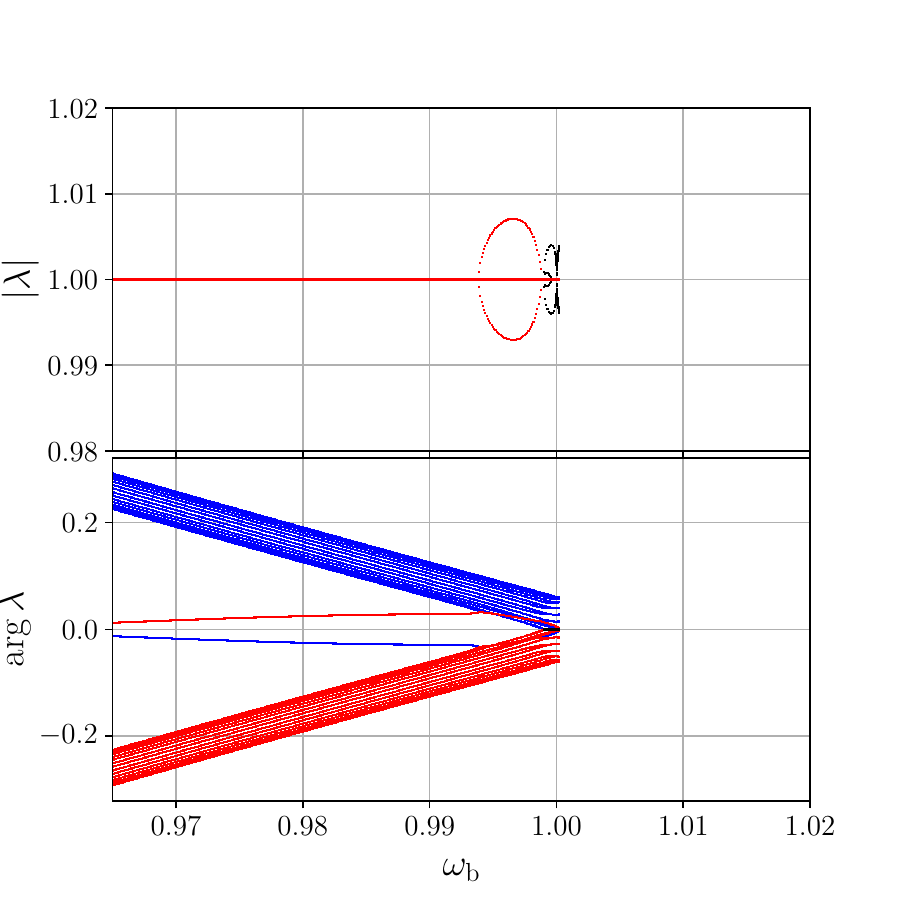}
\end{center}
\caption{Equal-period breathers: $\wb=\Omega$: Evolution of the moduli and angle of the Floquet multipliers with the breather frequency.
The sign of the Krein signature of the eigenvalue is represented by the code: blue $+$; red  $-$; and black as zero.
(Left) Breather in phase with the modulation; (Right) Breather in quadrature with the modulation.
The in-phase breather (left) is unstable with the unstable eigenvalues in the positive real line. However, it can be continued until the two bands of phonon eigenvalues collide. The in-quadrature breather (right) is stable and structurally stable as the close eigenvalues have the same Krein signature and cannot abandon the unit circle. Eventually, the isolated eigenvalues corresponding to the internal modes of the breather traverse the phonon bands with different Krein signatures bringing about oscillatory instabilities. The breather can be continued inside the phonon band.
Parameters: $\kappa=0.0052$, $\delta_1=0.1067$, $\delta_2=0.6996$, $N=16$.
}
\label{fig_floquetbr_evolution}
\end{figure*}

\begin{figure*}[p]
\begin{center}
\includegraphics[width=0.45\textwidth, clip, trim=0 8 0 40]{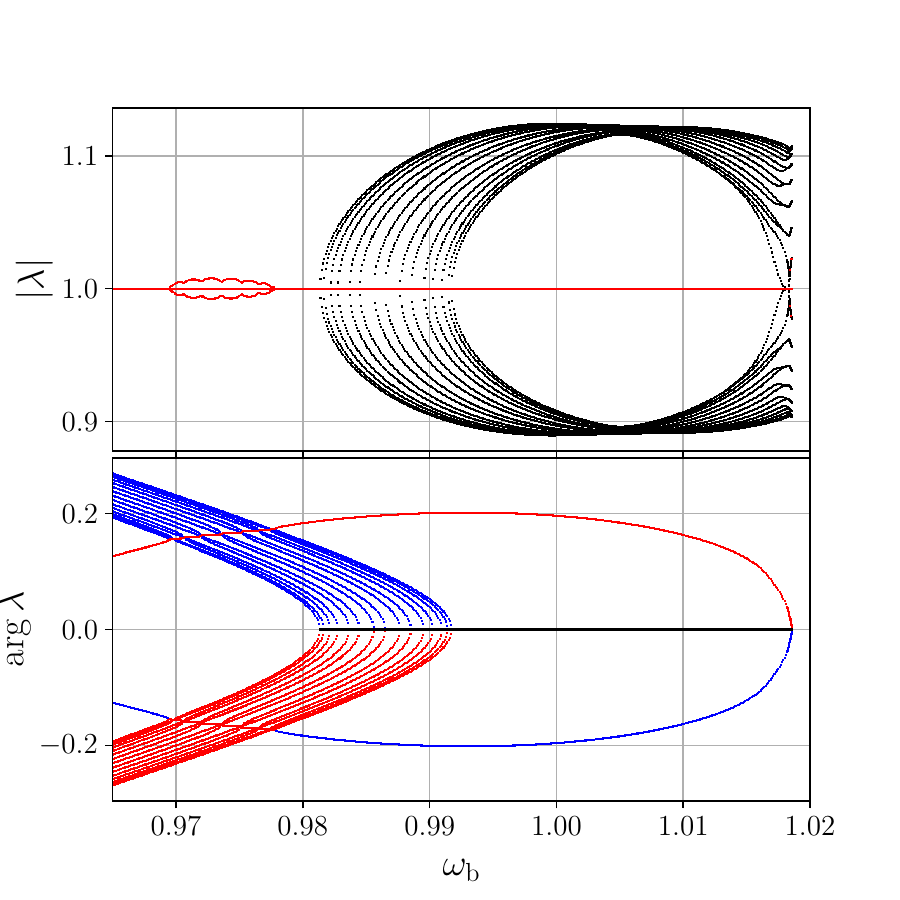}\mbox{}
\includegraphics[width=0.45\textwidth, clip, trim=0 8 0 40]{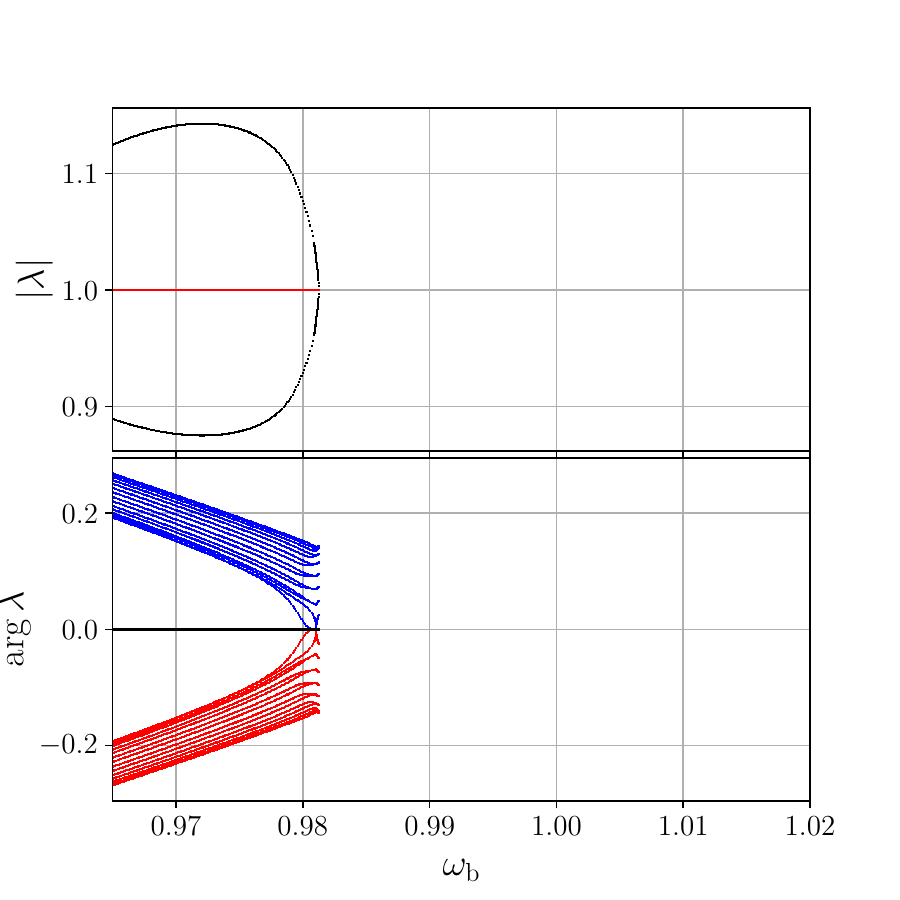}
\end{center}
\caption{Period-doubled breathers: $\wb=\Omega/2$: Evolution of the moduli and angle of the Floquet multipliers with the breather frequency.
The sign of the Krein signature of the eigenvalue is represented by the code: blue $+$; red $-$; and black as zero.
(Left) Breather in phase with the modulation; (Right) Breather in quadrature with the modulation.
The in-phase breather (left) is stable and structurally stable until  the isolated eigenvalues corresponding to internal modes of the breather traverse the phonon bands where oscillatory instabilities occur, and becomes again stable after going out of the phonon band. The crossing is apparent as the phonons have actually a $2\pi$ extra angle. It can be continued briefly until the two phonon bands collide bringing about instabilities at $+1$ as the system itself becomes unstable due to the resonance between the positive and negative phonon bands.  In spite of that, the breather can be continued.
The in-quadrature period-doubled breather (right) is unstable for all frequencies, with unstable multipliers in the positive real line.
Parameters: $\kappa=0.0052$, $\delta_1=0.1067$, $\delta_2=0.6996$, $N=16$.
}
\label{fig_doubled_br_evolution}
\end{figure*}

\section{Dependence of the breather stability with the frequency}
\label{sec:frequencydependence}
In this section, we observe the changes of stability and instability when both the breather frequency $\wb$ and $\Omega$ change simultaneously as they are commensurate as deduced before in Sec.\,\ref{sec:theory_ILM}.  The number of possible cases, i.e., depending on the fraction $\wb/\Omega$ is huge, and the stability changes for each case are difficult to analyze. A panorama of the many cases is shown in Fig.\,\ref{fig_phonon_band_breathers}, where a large number of cases for $\wb=(m_\text{b}/m_\text{m})\Omega$, with $m_\text{b}=10$ and $m_\text{m}=[2,\dots, 30]$ integers are presented indicating their stability or instability type. Note that those integers are not necessarily the same as in Lemma 1 and 2 in Sec.\,\ref{sec:exactmoving}, but their ratio is the same.

Each breather calculation is represented by a circle with a color coding its instability as indicated by the legend. The breather circles make apparent lines with constant slope $\wb/\Omega=m_\text{b}/m_\text{n}$. A particularly interesting case are equal-period breathers with $\wb=\Omega$, where it can be seen that they can be continued into the phonon band. Other case of interest are period-doubled breathers, $\wb=\Omega/2$, for which there is a resonance with the negative phonon band $\omega_q^{(-)}=-\omega_q=-\sqrt{\wo^2+2\kappa(1-\cos(q))}$, which is equivalent in a system with modulation frequency $\Omega$ to $\Omega-\omega_q$ and the two phonon bands collide. As there is no dissipation, the phonons take energy from the driving, grow and become nonlinear with their frequencies separating from the unmodulated phonon band.

When observing that a breather is unstable, that may be due to interaction with phonons and other causes. A frequent one is that there is another breather with different structure which is stable for the same parameters. We tested the existence and stability of breathers {\em in quadrature}, that is, with a phase difference of $\pi/2$ with the modulating term, with interesting results. Often, but by no means always, there is an interchange of stability between the in-phase breathers and the in-quadrature breathers as can be seen in Fig.\,\ref{fig_phonon_band_breathers}.

In this section, we concentrate into three cases of interest for ST breathers (i) equal-period breathers, with $\wb=\Omega$, period-doubled breathers $\wb=\Omega/2$, and period-halving breathers $\wb=2\Omega$. All of them have been found experimentally in photonic systems\,\cite{mukherjee-rechtsman2023,mukherjee-science2020,kang-period-halving2025} and the comparison may be of interest.

In order to present the stability evolution, a very useful method is the simultaneous representation of the moduli and arguments of the Floquet multipliers $|\lambda_i|$ and $\arg{\lambda_i}$. Usually, the existence of a breather  appears as a pair of isolated eigenvalues corresponding to internal modes of the breather, and the phonons as a group or band of eigenvalues, and, of course, this simple picture can change a lot. We think it is more practical to defer most of the explanation to the figure captions and give here some brief indications.

For equal-period Floquet breathers, the in-phase breather is unstable and the in-quadrature breather is stable, although the second one becomes unstable when its frequency collides with the phonon band as shown in Fig.\,\ref{fig_floquetbr_evolution}.

For period-doubled breathers the opposite happens: the in-phase breather is stable becoming unstable when crossing the phonon band but becoming stable again until the system itself becomes unstable as the two phonon bands collide. The in-quadrature breather is unstable but can be continued until the collision of the phonon bands. These properties can be seen in Fig.\,\ref{fig_doubled_br_evolution}.

For period-halved breathers, the in-quadrature breather is again stable, and the in-phase ones are unstable. The plots in this cases are trivial and not represented.

\section{Conclusions}
\label{sec:conclusions}
In this work we have studied the existence and stability of Floquet breathers in a space-time modulated nonlinear lattice. The physical system under study is a variation of a previously presented design of a cantilever array, with an on-site potential provided by magnetic forces created by electromagnets, which are now driven by DC and AC currents to create a modulation both in time and space.

The nonlinear dynamical equations are derived, and its linear limit is discussed. The dispersion bands, and their changes due to space, time, and space-time modulation have been obtained analytically and numerically.

We have adapted the theory of exact moving breathers to space-time modulated systems, concluding that the modulation and breather frequencies in the moving frame should be commensurate as a necessary condition for breather existence. For stationary breathers, the condition applies to laboratory frequencies.

The peculiarities of the stability analysis for time-modulated systems have have also been analyzed with the construction of an extended symplectic system.

First, breathers with a frequency multiple of the modulating one are studied and their properties analyzed. For modulating frequencies smaller than the breather frequency, single breathers and double breathers are found, both symmetric and anti-symmetric. When the modulating frequency becomes small, a phonobreather with extended background appears.

The evolution of the stability or instability as the frequency changes is analyzed in detail for in-phase and in-quadrature breathers, observing that for some cases they interchange their stability. Attention was paid to equal-period, doubled-period and halved-period Floquet breathers, that appear in photonic systems, observing the evolution of the Floquet eigenvalues with the frequency, obtaining insight of the phenomena.

Some theoretical results on moving breathers in modulated systems presented in this article will be numerically studied and checked in a forthcoming article.
This study reveals that this relatively simple system provides a huge variety of Floquet breathers which may also appear in other space-time modulated materials also known as dynamic metamaterials.

\section*{Acknowledgements}
JFRA acknowledges the Laboratory of Microdynamics at the University of Osaka for hospitality.

All the authors would like to acknowledge Serge Aubry for the inspiration of his work.
JFRA, in particular, although he never worked directly with him, he did so with close collaborators as Robert Mackay and Jos\'e Luis Mar\'in in Cambridge, in 1997, where discussions with them over their common publications with Aubry became his introduction to research in this field.

\section*{Funding}
MK acknowledges support from JSPS Kakenhi (C) No. 24K07393 and 21K03935.

JFRA and VJSM thanks grant PID2022-138321NB-C22 funded by MICIU/AEI/ 10.13039/501100011033 and ERDF/EU.

YD acknowledges support from JSPS Kakenhi (C) No. 24K14978.

\appendix
\renewcommand{\thesubsection}{\thesection.\arabic{subsection}}

\section{Numerical experiments, thermalization and noisy systems}
\label{app:thermal}

Along several parts of this work and in the sections below, we use the terms thermalization, near thermalization, or noisy systems. In this appendix we specify what we understand by those related concepts.

We perform numerical simulations and analytical calculations to obtain properties of the system as the initial coordinates and momenta for breathers, or the discrete Fourier components, or to find their stability properties. These simu\-lations are supposed to produce a definite and precise outcome. The imprecision is due to shortcomings of the numerical method, the parameters chosen, as time integration steps, the number of Fourier components, and so on.

We also perform {\em numerical experiments} to observe the system properties, similarly to physical experiments. The initial conditions are chosen at random with some desired energy or averages, and we observe the system evolve and  its route to an {\em approximate thermal equilibrium}, and its behavior once this thermal equilibrium is attained. An important example is to obtain the {\em numerically experimental} dispersion relation with different modulations as explained in Appendix\,\ref{app:dispersion}.

The term {\em approximate} appears because a rigorous definition of thermal equilibrium consists of the property that the statistical ensemble compatible with the macroscopic constrains allows the system to realize all the microscopic states with equal probability.\cite{reif2009} This is not achievable in practice, and the use of the ergodic theorem that a single realization of the ensemble also occupy all the microstates with equal probability if allowed infinite time is also difficult in practice.

Therefore, we consider that the system is in thermal equilibrium when some of its consequences are approximately met.
We understand that (i) the system forgets its microscopic initial conditions, i.e., positions and velocities but not energy and other quantities; (ii) its local energy probability distribution approaches the Boltzmann distribution; (iii) identical particles have the same properties over time; (iv) the participation function approaches $N/2$ (see below); (v) the site of the particle with peak local energy moves along the lattice, perhaps with  space-time correlation of some time when fluctuations bring about wave groups with some lifetime (See. Fig.\,\ref{fig_initial}) but not permanently unless they are a consequence of the system properties. Some temporary correlations are also properties of the system as phonon wave groups. Those are detected to construct, for example, the {\em numerically experimental} dispersion relations as explained in App.\,\ref{app:numericaldispersion_obtention}.

This subject is dealt with in Ref.\,\onlinecite{archilla2025} in the following way. Let us consider a system of particles with weak coupling in the microcanonical ensemble, we can suppose that each particle is in contact with a thermal bath corresponding to the rest of the system at a given temperature $T$, or thermodynamic $\beta=1/\mu$, with $\mu=k T$, $k$ being the Boltzmann constant. Within this approximation, the probability of a particle having an energy $e_n$ is proportional to $\exp(-\beta e_n)$,\cite{reif2009}  with $\mu$ equal to twice the average kinetic energy, according to the virial theorem.
\begin{figure}[t]
\begin{center}
\includegraphics[width=\largefigure]{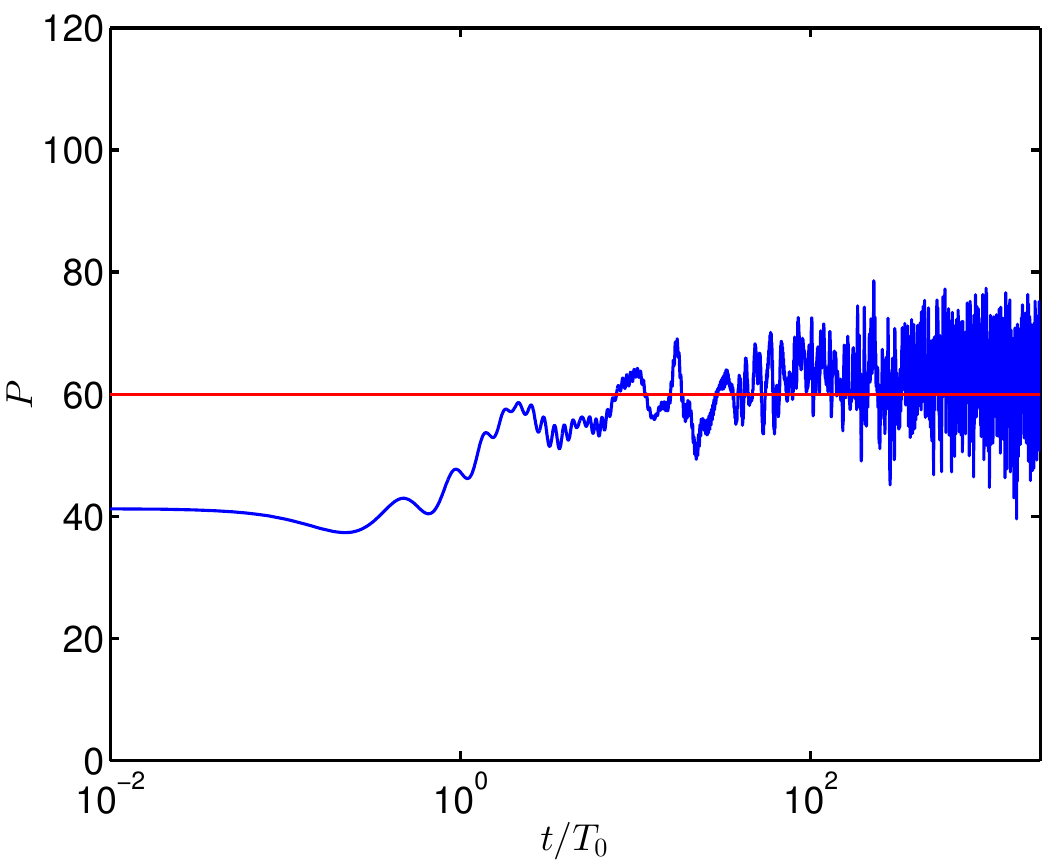}
\includegraphics[width=\largefigure]{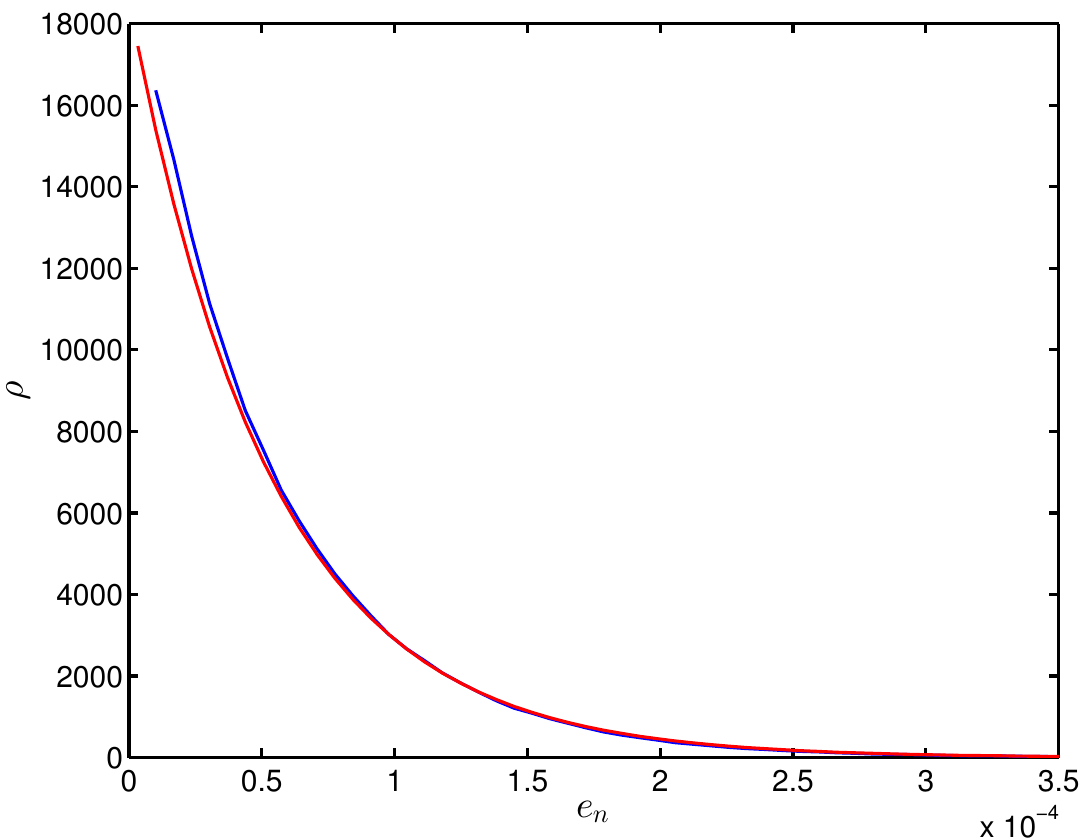}
\end{center}
\caption{({\bf Top:}) Path to thermal equilibrium for a space-modulated system with $h=2\pi/3$. The initial conditions are a random distribution of velocities with the coordinates at rest. The theoretical value at equilibrium is $P=N/2$ (in red), but with large fluctuations for a $N=120$ system. The logarithmic scale for time allows the observation both at the short scale at the beginning but also for longer times; ({\bf Bottom:}) Numerically experimental distribution function (blue) compared with the theoretical one (red). The first bin is discarded in the experimental curve. Parameters: $N=120$, $\kappa=0.10$, $\delta=0.05$, $\Omega=0$, $h=2\pi/3$.
  }
\label{fig_thermalspace}
\end{figure}

Imposing the probability one for a particle to have all the energies from 0 to $\infty$, the probability density of a particle having energy $e_n$ is $\rho(e_n)=\exp(-e_n/\mu)/\mu$, also known as the exponential distribution.
The average particle energy is $\langle e_n\rangle=\mu$, with standard deviation also $\mu$.
Averages are in principle for the statistical ensemble, but due to the ergodic theorem, the average can be made in time and also at the different identical particles.
\begin{figure}[t]
\begin{center}
\includegraphics[width=\largefigure]{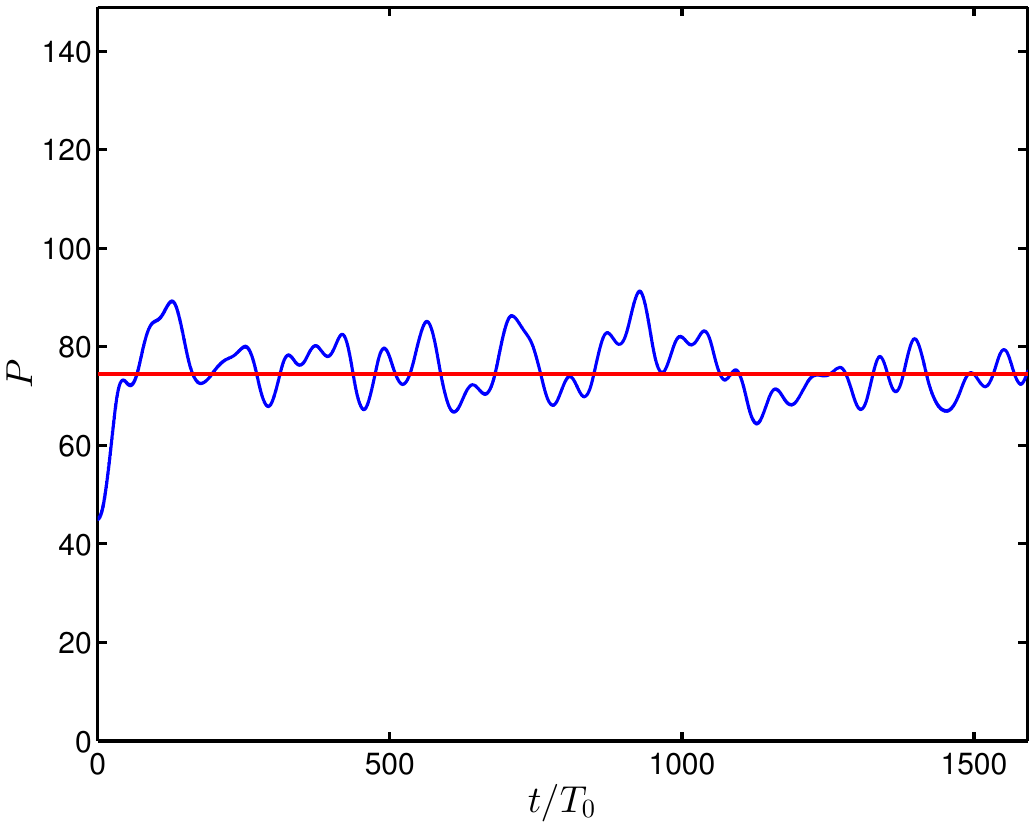}
\includegraphics[width=\largefigure]{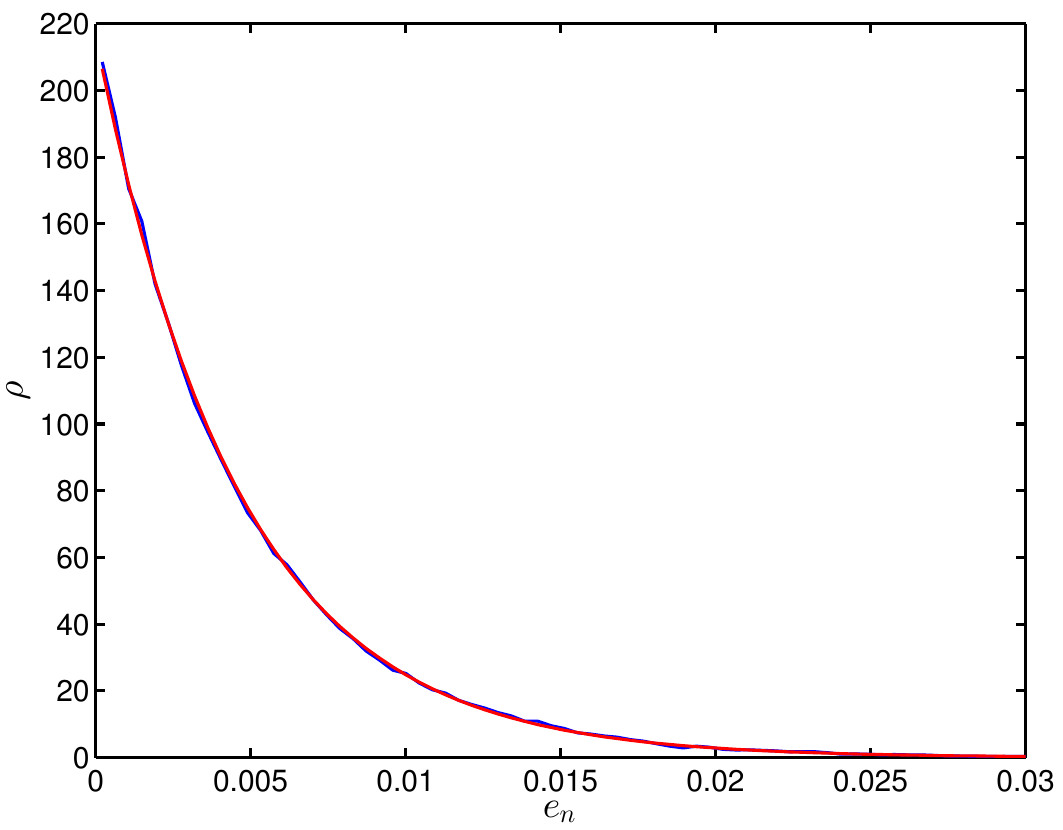}
\end{center}
\caption{({\bf Top:}) Path to thermal equilibrium for the particles outside a breather core. Initial conditions are a random distribution of velocities with the coordinates at rest except for the initial breather coordinates at a core of 11 particles in a system with $N=160$ particles. The theoretical value at equilibrium is $P=N/2$ (in red), but with large fluctuations for a $N=160-11=149$ system. In this case, a shorter time than in Fig.\,\ref{fig_thermalspace} is shown on a linear scale. ({\bf Bottom:}) Numerical experimental distribution function (blue) compared with the theoretical one (red)
outside the breather core.  Parameters: $N=160$, $\kappa=0.0052$, $\delta_1=0.1067$, $\delta_2=0.6996$, $\Omega=\wb/20=0.049$, $h=0$.
  }
\label{fig_thermalbreather}
\end{figure}

An easily measurable consequence of this distribution is the behavior of the participation function $P=(\sum e_n)^2/\sum e_n^2$. This is a function with value 1 if all the energy is located at a single particle and $N$ if the energy is evenly distributed among the system number of particles $N$. The value of $P$ at thermal equilibrium using the exponential distribution is $N/2$, with large oscillations of $P$ due to the large fluctuations of $e_n$ for the exponential distribution. The amplitude of the oscillation of $P$ around $N/2$ depends on the system size, the larger the size, the relative value of the oscillation becomes smaller as it is of the order of $1/\sqrt{N}$. As the distribution $\rho(e_n)$ is not exact, the actual value of the average $P$ can be a few particles above or below $N/2$, also changing with the time observation window, due to the large fluctuations of a small system with $N\simeq 100$.

We can check these properties for our system with different types of modulation. The exponential distribution fails at low energies when the model of a particle with energy $e_n$ interacting with the nearest neighbors is not correct as part of the energy is precisely the interaction energy with the nearest neighbors. Often the first bin to obtain the numerical distribution has to be discarded.

As an example, let us consider the simulations to  obtain the {\em experimental} dispersion relation for the system modulated in space with $h=2\pi/3$ with the theoretical and numerical results presented in Appendix\,\ref{app:onlyspacededuction}. The path to thermal equilibrium and the distribution function are represented in Fig.\,\ref{fig_thermalspace}.

\mbox{}\\
\noindent{\bf Noisy systems}\label{rev22-12}\\
The concept of a {\em noisy system} appears when we study breathers. The linear stability of breathers, and its nonlinear stability for small perturbations, is analyzed in Sec.\,\ref{sec:stability}, but another complementary technique of particular physical interest is to observe the breather behavior in presence of temperature. A system with a breather {\em is not thermalized}, because eventually the breather will disappear but the breather life can be very long for some systems and breather energies.\cite{archilla2025} A {\em noisy system} is a system that it is in an approximate thermal equilibrium, in the sense explained above, outside the breather core. The breather core is a few particles around the breather center, depending on the breather localization.

We add random velocities with some convenient kinetic energy to the initial coordinates for a system with a breather and perform numerical experiments with similar outcomes. The participation function of the whole system will be a few particles, but outside the breather core the same properties at thermal equilibrium considered above can be observed. This is illustrated in Fig.\,\ref{fig_thermalbreather}.


\section{Notation}
\label{app:notation}
To present  formally the theory, we will use the ket and bra notation. The nice thing of that notation is that the properties of operations with vectors, change of basis, and operators appear as the simple juxtaposition of kets and bras. We present here this notation; more details can be read in Ref.\,\onlinecite{griffithselectro2024} and in Ref.\,\onlinecite{archilla-multibreathers2003} applied to breather theory.

For our purposes, let as consider initially a ket $\ket{\phi}$ as a vector which can be expressed in different bases. There are two bases of interest in the domain of $H_0$, the first one is given by $\ket{n}'s$, with $n=0:N-1$, acting on any eigenfunction $\ket{u}$ of $H_0$ as $\braket{n}{u}=u_n$. In that same basis is the column matrix $\ket{n}=[0,0,\dots, 0, 1,0,\dots,0]'$, with the only $1$ in position $n$ and the rest zeros. A vector of positions are $\ket{u}=[u_1,u_2,\dots,u_n]'$ can be seen as $\ket{u}=\sum_{n=0}^{N-1} u_n\ket{n}$.

The bra $\bra{\phi}$ is the Hermitian conjugate, so $\bra{\phi}$ is a row matrix with elements the complex conjugates of the elements of $\ket{\phi}$.
In this way, a bracket $\braket{\phi_1}{\phi_2}$ is the scalar product $\sum_{n=0}^{N-1}\phi_{1,n}^*\phi_{2,n}$. Trivially $\braket{n}{n'}=\delta_{n,n'}$. So the basis $\{\ket{n}\}$ is orthonormal. $\braket{n}{\phi}=\phi_n$ is the scalar product by an unit basis vector and, therefore, the component of $\ket{\phi}$ on that basis vector.

A dyad $\dyad{\phi_1}{\phi_2}$ is a linear operator acting on kets as $\dyad{\phi_1}{\phi_2}\ket{\phi_3}=\braket{\phi_2}{\phi_3}\ket{\phi_1}$

Any linear operator $H$ can be expressed in a given basis as a sum of dyads as  $H=\sum_{n=0}^{N-1}\sum_{n'=0}^{N-1} H_{n,n'}\dyad{n}{n'}$, with $H_{n,n'}=\bra{n}H\ket{n'}$ being the elements of the matrix representation of $H$ in the basis $\{\ket{n}\}$.

The second basis is the momenta basis, and it is given in the  $\{\ket{n}\}$ basis as:
\begin{align}
\ket{q_k}&=\ket{k}=\fracc{1}{\sqrt{N}}\sum_{n=0}^{N-1}\exp(-\ii\frac{2\pi k n}{N})\ket{n}, \quad \mathrm{then}\quad \nonumber\\
\ket{n}&=\frac{1}{\sqrt{N}}\sum_{k=0}^{N-1}\exp(\ii\frac{2\pi k n}{N})\ket{k}.
\end{align}
The second expression is the representation of $\ket{n}$ in the $\ket{k}$-basis.

The following relations are useful and easy to obtain:
\begin{align}
\braket{n}{n'}&=\delta_{n,n'} \quad ;\quad \braket{k}{k'}=\delta_{k,k'}\quad ; \\
\braket{k}{n}&=\fracc{1}{\sqrt{N}}\exp(\ii\frac{ 2\pi k n}{N})\quad ;\quad \braket{n}{k}=\fracc{1}{\sqrt{N}}\exp(-\ii\fracc{2\pi n k}{N})
\end{align}

The identity operator is given by $I=\sum_{n=0}^{N-1}\dyad{n}{n}=\sum_{k=0}^{N-1}\dyad{k}{k}$, then
\begin{align}
\ket{u}&=\sum_{n=0}^{N-1}\ket{n}\braket{n}{u}=\sum_{n=0}^{N-1}u_n\ket{n}\, , \quad \mathrm{and}\quad\nonumber\\
\ket{u}&=\sum_{k=0}^{N-1}\ket{k}\braket{k}{u}=\sum_{k=0}^{N-1}F_k(u)\ket{k},
\end{align}
where $\braket{n}{u}=u_n$, is the $n$-component of $\ket{u}$ in the $\{\ket{n}\}$ basis,  and
$\braket{k}{u}=F_k(u)$ is the $k$-component of $\ket{u}$ in the $\ket{q_k}$ basis.

Using the appropriate form of the identity operator we can obtain the relationship between $u_n$ and $F_k(u)$, that is
the inverse and direct discrete Fourier transforms.
\begin{align}
u_n&=\braket{n}{u}=\bra{n}\left(\sum_{k=0}^{N-1}\dyad{k}{k}\right)\ket{u}=\sum_{k=0}^{N-1}\braket{n}{k}F_k(u)\nonumber\\
&=\sum_{k=0}^{N-1}\frac{1}{\sqrt{N}}\exp(-\ii\fracc{2\pi n k}{N})F_k(u)\, , \\
F_k(u)&=\braket{k}{u}= \bra{k}\left(\sum_{n=0}^{N-1}\dyad{n}{n}\right)\ket{u}\nonumber\\
&=\sum_{n=0}^{N-1}\braket{k}{n}\braket{n}{u}=
\sum_{n=0}^{N-1}\fracc{1}{\sqrt{N}}\exp(\ii\fracc{2\pi n k}{N})u_n
\end{align}

\subsection{Functions of position and time}
\label{app:func_pos_time}
Suppose we have
 a function $u$ of the particle number $n$ and time $t$, its value at a particle and time $u_n(t)$, are  the components of the function $u$ represented as the ket $\ket{u}$ in the basis with elements $\ket{n,t}$, i.e. $\braket{n,t}{u}=u_n(t)$. In principle, $t$ is a continuous variable but, in practice, it is a discrete one obtained by the sampling of numerical integration. So, if there are $N_t$ time samples, $t_l=l\Delta t$, with $l=0:N_t-1$ and $\Delta t$ the sampling interval.  We change the basis notation to $\ket{n,l}$, with $\braket{n,t}{u}=u_n(t_l)=u_n(l\Delta t)$
and
$$\ket{u}=\sum_{n=1}^{N}\sum_{l=0}^{N_t-1} u_n(t_l)\ket{n,l}\, .$$
The basis $\ket{n,l}$ is orthonormal, i.e. $\braket{n,l}{n',l'}=\delta_{n,n';l,l'}$.

We assume that $u_n(t_l+N_t)=u_n(t_l)$, i.e., that $u$ is time periodic with period $T_f=N_t\Delta t$. This is not a limitations as $T_f$ is large enough that it has no consequences, similarly to Born-Von Karman boundary conditions in space.

Then, the alternative basis is given by the harmonic functions $\ket{k,s}$, with components:
\begin{align}
\braket{n,l}{k,s}=\fracc{1}{\sqrt{N N_t}}\exp(-\ii [q_k n-\omega_s t_l])\, ,
\end{align}
with period $T_f$, i.e., $\omega_s=\fracc{2\pi s}{T_f}=\fracc{2\pi s}{N_t\Delta t}$, ($s=0\dots N_t-1$) with  maximum frequency $~\fracc{2\pi}{\Delta t}$. Usually, frequencies are shifted so as the zero frequency is the middle and then the maximum frequency is the Nyquist frequency  $2\pi/2\Delta t$. Note that $\Delta t$ determines the largest frequency, but the resolution is given by $\fracc{2\pi}{T_f}$, that is, it depends on the final time.

The basis $\ket{k,s}$ is also orthonormal as
\begin{align} \braket{k,s}{k',s'}=\delta_{k,k';s,s'}\, .
\end{align}
Therefore,
 \begin{align}
 \braket{k,s}{u}&=\sum_{k=0}^{N-1}\sum_{s=0}^{N_t-1}\braket{k,s}{n,l}\braket{n,l}{u}\\
 &=\fracc{1}{\sqrt{N N_t}}\sum_{k=0}^{N-1}\sum_{s=0}^{N_t-1}\exp(-\ii [q_k n-\omega_s t_l]) u_n(t_l)
 \end{align}

We can also obtain  the expression of $\fracc{\d^2}{\d t^2}$ in $k,w$ space. For that:
\begin{align}
\fracc{\d^2}{\d t^2}\ket{k',s'}=-\omega_{s'}^2\ket{k',s'}
\end{align}
and then:
\begin{align}
\bra{k,s}\fracc{\d^2}{\d t^2}\ket{k',s'}=-\omega_{s'}^2\delta_{k,k';s,s'}
\end{align}

\section{Derivation of the effect of space-time modulation on the linear dispersion}
\label{app:dispersion}
The dispersion relations obtained in numerical experiments and the theoretical ones have already been presented in
Sec.\,\ref{sec:dispersionrelation}, with the exception of only space modulation presented below as it was not essential for the rest of the article.  Here, after some short introduction, we
describe the method to visualize the dispersion relations in numerical experiments followed by the theoretical derivations in several cases.

We consider the linear system described above in \eqref{eq:linearscaled} which is valid for small displacements $u_n$.

For $\delta=0$, the system is a well known, Klein-Gordon system, having a basis of solutions $\exp(\ii q n-\omega t)$, with $\omega_q^2=\wo^2+2\kappa (1-\cos(q))$. Note that for a given $q$ there are two frequencies $\pm \omega$ that are also frequencies for $-q$.

For a finite system with $N$ units and periodic boundary conditions the possible values of $q$ are $q_m=2\pi m/N$, with $m=0,\dots, N-1$, this includes also negative values of the wavevector $q=-q'$, with $q'>0$, as any wave number $q$  is equivalent to $q\pm 2\pi n$. Also, any set of numbers $m'=m+n$, where $n$ is any integer, leads to an equivalent basis. A convenient one, centered around $q=0$ is $-N/2,\dots, N/2-1$ if $N$ is even, or $-(N-1)/2, \dots, (N-1)/2$ if $N$ is odd. An alternative basis in real space is given by $\cos(q n-\omega t)$ and $\sin(q n-\omega t)$, but the exponential functions basis is more convenient analytically.\label{rev22-16}

\subsection{Obtention of the dispersion curves in numerical experiments}
\label{app:numericaldispersion_obtention}
We obtain numerically what we could call the experimental dispersion curves in an approximate thermal equilibrium, as explained with detail in App.\,\ref{app:thermal},
with the following procedure: First we consider a relatively large system, with, for example, $N=128$ oscillators, and use as initial conditions a random distribution  of positions for  the variables $u_n$ in [-0.5,0.5],
Equally, the random perturbation can be added to the momenta or both the momenta and coordinates without changing the outcome.
Second, we leave the linear system \eqref{eq:linearscaled} to evolve a relatively long time, $t=5000$, for example, when the system is supposed to be near or at thermal equilibrium and, therefore, all the linear modes or phonons are in principle excited. We can also use the full nonlinear system but with conveniently small values of the coordinates or momenta so that the system is near the linear limit.
Then, we use the final positions and velocities, to run a simulation of similar length. We use the results to perform the two-dimensional discrete Fourier transform in space and time (XTDFT) on the variable $u_n(t)$, that is,
\begin{align}
F(q_k,\omega_r)=\frac{1}{\sqrt{N_t N}}\sum_{n=0}^{N-1}\sum_{l=0}^{N_t-1}\exp(\ii[q_k n-\omega_r t_l])u_n(t_l)\, ,\nonumber
\label{eq:XTDFTdef}
\end{align}
with $t_l=\fracc{l}{N_t}t_f$, $\omega_r=\fracc{r}{N_t} \fracc{2\pi}{\Delta t}$, for $l,r=0\dots N_t-1$
and $q_k=\fracc{k}{N}2\pi$, for $k=0\dots N-1$.
The sampling interval $\Delta t=\fracc{t_f}{N_t}$ is larger than the integration step to avoid very large matrices as the maximum frequency $2\pi/\Delta t$ does not need to be very large because the frequencies are of the order of $\omega_0$ and $\Omega$. On the other hand the frequency precision $\Delta \omega=\fracc{2\pi}{N_t\Delta t}=\fracc{2\pi}{t_f}$,  is inversely proportional to $t_f$, therefore, a large value of it is convenient, both to obtain precision for the frequency and for the results to be generic enough.

The intensities are defined as $I(\omega_i,q_n)=|F(\omega_i,q_n)|^2$, normalized to unity. To visualize the dispersion curves,  we represent a number of lines in contour plots, typically ten, to get a clear picture. However, all the modes do not have the same probability to occur, the density of states may be small, and theoretically existing modes may have small intensities and are difficult to discriminate from the noise background, those dispersion lines are {\em weak}. Presently, we do not have a means to calculate those intensities theoretically.

This method is changed to check that the results do not depends on the details,  the random distribution can be applied only to momenta or to both coordinates and momenta; we can change the thermalization time, simulation time, sampling time interval, number of particles, etc. This method can also be used with a superposed breather in other sections.

The result of the simulation after thermalization are also used to obtain energy density plots. An example of both is shown in Fig.\,\ref{fig_initial}.
We can see that new dispersion bands appear roughly parallel to the unmodulated one and with their maxima and minima displaced.

\subsection{Dispersion bands modification with only space modulation}
\label{app:onlyspacededuction}
In order to understand the results, we  will consider first the case $h=2\pi/L$, where $L$ is an integer, and $\Omega=0$. In this case, the symmetry of the system $n\rightarrow n+1$ is broken and substituted by the symmetry $n\rightarrow n+L$.
Let us suppose that $\delta=0$, but still the symmetry is broken, then, in order to be able to impose periodic boundary conditions, we need that $N=RL$ with $R$ and integer number. Now, by Bloch theorem,\cite{griffithsquantum2018}
the solutions are of the form $\exp(\ii q n)F(n,t)$, with $F(n,t)$ a solution with the periodicity of the lattice, whose period is now $L$, that is the elements of the basis are given by $u_{l,r}=\exp(\ii q_r n) \exp[\ii(q_l n-\omega t])$. As the second factor has the periodicity of the new lattice $q_l=(2\pi/L) l$, with $l=0, \dots, L-1$. If we impose periodic boundary conditions, then $u_{l,r}(n,t)=u_{l,r}(n+N,t)$, the second factor is automatically identical if $N=RL$ and the first factor becomes $q_r=2\pi r/N$, with  $r=0, \dots, R-1$, as larger values of $r$ can be written as $q_r'=2\pi (r_1 R+r_2)/N=2\pi (r_1 R+r_2)/LR=2\pi r_1/L +2\pi r_2/N$, and the first term is periodic in $L$ and can be incorporated into $F(n,t)$. We can rewrite it as:
\begin{align}
u_{n,t}&=\exp(\ii q_l n)\exp(\ii q_r n)\exp(-\ii \omega_{l,r} t) \\ 
&=\exp(\ii\frac{2\pi r}{N}n)\exp(\ii\frac{2\pi l}{L}n)\exp(-\ii\omega_{l,r}t)\, ,
\label{eq:unbroken}
\end{align}
with $r=0,\dots, R-1$ and $l=0\dots L-1$. The original index $m=0, \dots\, N-1$ is given by $m=rL+l$, and the frequencies are given by
\begin{align}
\omega_{l,r}^2=\wo^2 + 2\kappa(1-\cos(\frac{2\pi l}{L}+\frac{2\pi r}{N}))\, .
\label{eq_wphon_broken2}
\end{align}

In the following, we use the standard notation of ket and bras summarized in Appendix\,\ref{app:notation}.

The solutions of the system $H_0 \ket{u}=E_{0}\ket{u}$, with $E=\omega^2$ are given by the eigenvectors of $H_0$ with the same eigenvalue $E=\omega^2$. There are two eigenvectors with the same eigenvalue, corresponding to
$\ket{k}$ and $\ket{-k}$, the latter being equivalent to $\ket{N-k}$ under periodic boundary conditions. Their eigenvalues are $E_0=\omega^2=\wo^2+2\kappa(1-\cos(\fracc{2\pi k}{N}))$.

The solution to the eigenvalue equation of the perturbed Hamiltonian
$$(H_0+\delta H_1)\ket{\phi}= (E_0+\delta E_1)\ket{\phi}
$$
will be in the first approximation close to the eigenspace of $E_0$, that is some linear combination of $\ket{k_1}\equiv\ket{k}$ and $\ket{k_2}\equiv\ket{-k}$ (it can be done with the cosine and sine solutions but the theory has to be reformulated and does not add any value). So, let us suppose that $\ket{\phi_1}$ is a
a linear combination of the orthogonal eigenvectors, i.e., $\ket{\phi_1}=\alpha\ket{k_1}+\beta\ket{k_2}$ and then, we obtain
$$
E_0\ket{\phi}=H_0\ket{\phi}\,\quad \mathrm{and}\quad  E_1\ket{\phi}=H_1 \ket{\phi},
$$
and we can write $\ket{\phi}=\ket{\phi_1}$.
The first equation holds because $\phi$ is in the $E_0$-eigenspace of $H_0$. Substituting $\ket{\phi_1}$ and using that $H_1$ is linear, we obtain:
$$
E_1(\alpha\ket{k_1} +\beta \ket{k_2})=\alpha H_1\ket{k_1}+\beta H_1 \ket{k_2}\, .
$$
Multiplying by the left by $\bra{k_1}$ and $\bra{k_2}$ and using that $\braket{k_1}{k_2}=\delta_{1,2}$, we obtain:
\begin{align}
\alpha E_1 &=\alpha\bra{k_1}H_1\ket{k_1}+\beta\bra{k_1}H_1\ket{k_2}\\
\beta E_1 &=\alpha\bra{k_2}H_1\ket{k_1}+\beta\bra{k_2}H_1\ket{k_2}
\end{align}

In other words and in general, for any number of degenerate orthonormal eigenvectors $\ket{k_i}$ with the same eigenvalue, the corrections $E_1$ to the initial eigenvalue $E_0$ are given by the eigenvalues of the matrix $M$, which is the matrix representation of $H_1$ in the $E_0$-subspace, or, explicitly
\begin{align}
M_{ij}=\bra{k_i}H_1\ket{k_j}\,
\end{align}
The eigenvectors of $M$ provide the linear combination of $\ket{k_i}$ with a eigenvalue $E_1^i$. By construction $M$ is hermitian $M_{ij}=M_{ji}^*$ and the eigenvalues $E_1^i$ are real.

What is left is to obtain $M$, which is done by
\begin{align}
\bra{k_i}H_1\ket{k_j}&=\bra{k_i}\left(\sum_{n=0}^{N-1}\dyad{n}{n}\right)H_1\left(\sum_{n=0}^{N-1}\dyad{n'}{n'}\right)\ket{k_j}\nonumber\\
&=\sum_{n=0}^{N-1}\sum_{n'=0}^{N-1}\braket{k_i}{n}\bra{n}H_1\ket{n'}\braket{n'}{k_j}
\end{align}

In our case $H_1\ket{n}=\cos(h n)\ket{n}$ and $\bra{n}H_1\ket{n'}=\cos(h n)\delta_{n,n'}$.  Therefore, we obtain:
\begin{align}
\bra{k_i}H_1\ket{k_j}=\frac{1}{N}\sum_{n=0}^{N-1}\exp(\ii\fracc{2\pi k_i n}{N})\cos(h n)\exp(-\ii\fracc{2\pi k_j n}{N})
\end{align}

If $h=2\pi m$, with $m$ integer,  then $\bra{k_i}H_1\ket{k_i}=1$, but it is a case of no interest, as the system would have again unit lattice distance (or less). Otherwise, if $h\neq 2\pi m$:
\begin{align}
\bra{k_i}H_1\ket{k_i}&=\fracc{1}{N}\sum_{n=0}^{N-1}\cos(hn)\, .
\end{align}
As it does not depend on $k_i$, the diagonal terms of $M$ are equal and real, given by:
\begin{align}
a=M(1,1)=M(2,2)=\Re\left(\fracc{1}{N}\sum_{n=0}^{N-1}\exp(\ii h n)\right)\, .
\label{eq:diagonal}
\end{align}
Let us call $b=M(1,2)=M(2,1)^*$, then, the eigenvalues of $M$ are $E_1=\Delta\omega^2=a\pm|b|$. The value $a$ represents
a global shift of the dispersion curve, while $b$ represents the separation in two values of $\omega^2$.

However, the term \eqref{eq:diagonal} is bounded by 1 at $h=0$, which is of no interest. It is also zero for $h=2\pi/L$ with $L=N/R$, $L$ and $R$ integers. For other values of $h>0$ it is very small except at the proximity of $h=0$.
Therefore, the global shift provided by the modulation is very small.


The non-diagonal elements of $M$ are complex conjugates and given by:
\begin{align}
b&=\bra{k}H_1\ket{-k}=\fracc{1}{N}\sum_{n=0}^{N-1}\exp(\ii 2 q_k n)\cos(hn)\nonumber \\
&=\fracc{1}{2N}\sum_{n=0}^{N-1}\exp(\ii[2q_k+h]n)+\fracc{1}{2N}\sum_{n=0}^{N-1}\exp(\ii[2q_k-h]n)\, .
\label{eq:bspread}
\end{align}
This expression is bounded by 1, taking its maximum absolute value 1 when both  $2q_k$ and $h$ are integer multiples of $2\pi$. But that case is of no interest at it implies $L=1$ with no symmetry breaking, so $h\leq \pi$.  However, $b$ is close to 0.5 if at least one of the two sums in \eqref{eq:bspread} is $N$, which happens for either $2q_k+h=2\pi m$ or $2q_k-h=2\pi m'$, for $m$ and $m'$ integers. This condition is fulfilled for $q_k\in [0 ,2\pi]$ at four  values, although they degenerate to a single value when $h=\pi$ as $-h/2=h/2 -\pi$.

If $h<\pi$:
\begin{align}
&b=M(1,2)=\bra{k}H_1\ket{-k}\simeq 0.5\,\quad \mathrm{for}\quad \nonumber\\
&q=\fracc{h}{2},\, q=\fracc{h}{2}+\pi,\, q=2\pi-\fracc{h}{2}, q=\pi-\fracc{h}{2}\, .
\end{align}

Therefore, the effect of the modulations is a small shift in $\omega^2(q)$ of order $\delta a$ and the degeneracy raising  of the bands for the four given specific wavenumbers is of or order $\delta/2$.

As an example, if $h=2\pi/3$, then $a=0$, and the values of $q$ for which the bands split are $\pi/3,\,2\pi/3,\,4\pi/3,\, 5\pi/3$.  Figure\,\ref{fig_phononL3} shows the numerically obtained dispersion function and the theoretical results on degeneracy raising.
Note that the dispersion curves are represented within the Brillouin zone before modulation $[0,2\pi]$. The Brillouin zone after modulation is $[0,2\pi/L]=[0,2\pi/3]$, with $L=3$ the new lattice distance after the breaking of the symmetry invariance. Phonon wavevectors $q>2\pi/3$ have to be translated to $q-n2\pi/3$, with n=1 or 2, to the new Brillouin zone. We think that this extended representation is clearer for the present context.

\begin{figure}[htbp]
\begin{center}
\includegraphics[width=\largefigure]{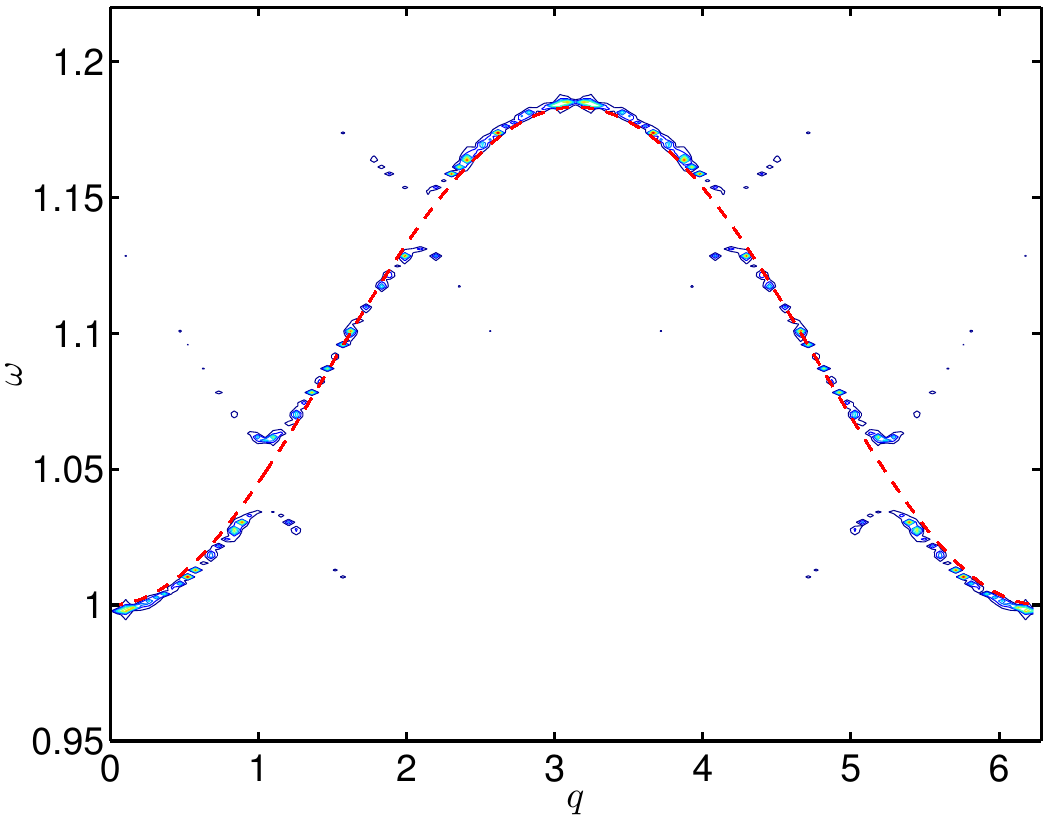}
\includegraphics[width=\largefigure]{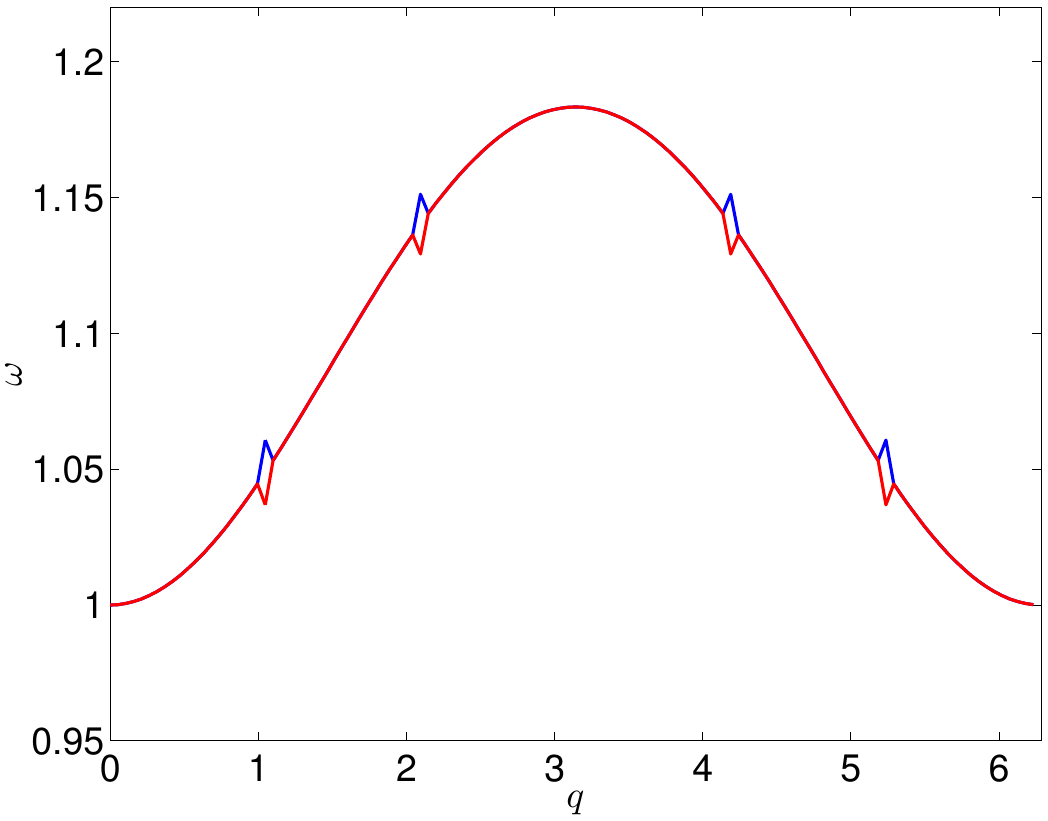}
\end{center}
\caption{({\bf Top:}) Numerical dispersion bands for $h=2\pi/3$; ({\bf Bottom:}) Theoretical degeneracy raising of the
dispersion bands due to space modulation. Note that the Brillouin zone is $[0,2\pi/3]$. See the text for the explanation.
  Parameters: $\kappa=0.10$,\label{rev22-17}
  $\delta=0.05$, $\Omega=0$ .
  }
\label{fig_phononL3}
\end{figure}



\subsection{Effect of only time modulation on the phonon bands}
\label{app:timemodulation}
When $h=0$, the effect of time modulation is to split the phonon band in several parallel bands separated by a frequency  $\Omega$.  They are obtained numerically as explained in
App.\,\ref{app:numericaldispersion_obtention}. Some bands are {\em weak}, meaning that their intensity in the XTDFT is small, presumably  because their density of states is small. Some become so weak that they cannot be seen into the background noise.
The number of bands diminishes with increased frequency and eventually the bands disappear or are not visible. Extra bands appear for frequency below $\omega_0=1$ although they are very weak when approaching $\Omega=\wo$.
Frequencies multiples of $\wo$ lead to the divergence of the system. Frequencies inside the phonon band excites the modes with that frequency.

These bands can be obtained analytically using the breakup of the time-invariance brought about by the time modulation. For that,
we will use the notation and theory treated
in Appendix\,\ref{app:notation}, taking into account the changes brought about by the time dependence of the on-site potential.

Consider the unmodulated version of Eq.\,\eqref{eq:linearscaled}
\begin{equation}
\ddot u_n=-\wo^2 u_n+\kappa(u_{n+1} + u_{n-1}-2 u_n)\, .
\label{eq:unmodulatedlinear}
\end{equation}
It is invariant under translations in time $t\rightarrow t+\Delta t$, where $\Delta t$ is the sampling interval, which could be arbitrary, in principle. Let us choose $\Delta t=T_m/M$, with $M$ an integer, that is $\Omega=2\pi /T_m=M 2\pi/\Delta t$. The addition of the term $-\delta\cos(\Omega t)u_n$ to the rhs of \eqref{eq:unmodulatedlinear}  breaks this invariance. The modulated system is invariant only under larger translation of time $T_m=M\Delta t$, multiples of $\Delta t$.
There is a breaking of the symmetry:  the system is no longer invariant under the time shifts $t\rightarrow t+\Delta t$ but only under a subgroup of time shifts $t\rightarrow t+T_m=t+M\Delta t$.

The new solution is given by Bloch theorem,\cite{slater1958,cassedy1963} as $\exp(-\ii\omega t)$, multiplied by a periodic function with the period of the operator, that is
$$
u=\exp(\ii[q n-\omega t])\sum_{m=-M/2}^{M/2} A_m\exp(-\ii m\Omega t)\, ,
$$
with the coefficients $A_m$ depending on $q$.

Then, the substitution of  $u$ into the dynamical system leads to (substituting $\cos(\Omega t)=(\exp(\ii \Omega t)+\exp(-\ii\Omega t))/2$:
\begin{align}
(\omega+m\Omega)^2 A_m & =\big[\wo^2+2\kappa (1-\cos(q))\big] A_m \nonumber \\\
&+\fracc{\delta}{2}(A_{m+1}+A_{m-1})\, ,
\end{align}
or
\begin{align}
(\omega+m\Omega)^2 A_m=\bar\omega_q^2 A_m+\fracc{\delta}{2}(A_{m+1}+A_{m-1})\, ,
\end{align}
where $\bar\omega_q$ is the unperturbed frequency for the wavenumber $q$.
%
This is a complicated equation, but if $\delta$ is small with respect to $\wo^2$, we can suppress that term in the first approximation, that is.
\begin{align}
(\omega+m\Omega)^2 A_m=\bar\omega_q^2 A_m\, .
\label{eq:timewsquared}
\end{align}

This equation has a solution corresponding to make all  $A_k=0$, except $A_m$, for each $m$, indicating that each solution is characterized by a harmonic of higher order of the modulating wave.

Then, we obtain:
\begin{align}
\omega_{q,m}=-m\Omega+\sqrt{\bar\omega_q^2}=-m\Omega+\sqrt{\wo^2+2\kappa(1-\cos(q))}\, .     
\end{align}
Therefore, the main effect of the symmetry breaking is the appearance
of bands separated by the frequency $\Omega$ almost parallel to the unperturbed dispersion band (UDB). However, the intensity of those bands tend to be smaller the further apart from the UDB. It is necessary to plot contours of small intensity  in order to observe them.

Solving \eqref{eq:timewsquared}, it is also possible to obtain $\omega=m\Omega-{\bar\omega}_q^2$, but this equation lead to a change of sign of the frequencies, corresponding to waves traveling in opposite direction, which are already obtained for negative wavenumbers $q$.

Two examples can be seen in Fig.\,\ref{fig_timemod}

\subsection{Dispersion bands for space-time modulation}
\label{app:space-time}
Considering the system \eqref{eq:linearscaled} with both $\Omega\neq0$ and $h\neq 0$:\label{rev22-18A}
\begin{equation}
\ddot u_n=-\wo^2u_n -\delta\cos(hn-\Omega t)u_n+\kappa(u_{n+1} + u_{n-1}-2 u_n)\, .
\label{eq:dynamicalspace-time}
\end{equation}
The system is now $2\pi$ periodic in the variable $hn-\Omega t$, and the solutions can be written as the product of a plane wave multiplied by
a function with the periodicity of the system.\cite{cassedy1963} That is:
\begin{align}
u_n(t)=\exp(\ii[q n-\omega t])\sum_{m=-K/2}^{{K}/2} A_m\exp(\ii m[h n-\Omega t])\, ,
\end{align}
with $K$ being an integer large enough to obtain the convergence of the truncated Fourier series.\label{rev22-18B}
Then, the substitution of  $u_n(t)$ in the Hamiltonian leads to (substituting $\cos(hn-\Omega t)=(\exp(\ii[hn- \Omega t])+\exp(-\ii[hn-\Omega t))/2$):
\begin{align}
(\omega+m\Omega)^2 A_m&=\big[\wo^2+2\kappa (1-\cos(q+m h))\big] A_m\nonumber\\
+&\fracc{\delta}{2}(A_{m+1}+A_{m-1})\, .
\end{align}
Assuming that $\delta$ is small with respect to $\omega_0^2$,\label{rev22-4} we can obtain the independent solutions where all the $A_k=0$, except one $A_m$ each time, so it corresponds to a single harmonic with $\omega$:
\begin{align}
\omega= -m\Omega +\sqrt{\wo^2+2\kappa (1-\cos(q+m h))}
\end{align}
An example can be seen in Fig.\,\ref{fig_mod_spt}.


\newcommand{\noopsort}[1]{} \newcommand{\printfirst}[2]{#1}
  \newcommand{\singleletter}[1]{#1} \newcommand{\switchargs}[2]{#2#1}

\end{document}